\documentclass[pra,aps,showpacs]{revtex4}
\usepackage{bm}
\usepackage{amsmath,amssymb}
\usepackage{graphicx}

\begin{document}

\title{Tensorial analysis of the long-range interaction between \\
metastable alkaline-earth atoms}
\author{Robin Santra}
\author{Chris H. Greene}
\affiliation{JILA, University of Colorado,
Boulder, CO 80309-0440}
\date{\today}
\begin{abstract}
Alkaline-earth atoms in their lowest 
$\left(n{\mathrm{s}}n{\mathrm{p}}\right)$~$\phantom{}^{3}P\phantom{}_{2}$ 
state are exceptionally long-lived and can be trapped magnetically.
The nonspherical atomic structure leads to anisotropic long-range interactions
between two metastable alkaline-earth atoms. The anisotropy affects the 
rotational motion of the diatomic system and couples states of different 
rotational quantum numbers. This paper develops a tensorial decomposition of 
the most important long-range interaction operators, and a systematic inclusion
of molecular rotations, in the presence of an external magnetic field. This 
analysis illuminates the nature of the coupling between the various 
degrees-of-freedom. The consequences are illustrated by application to a 
system of practical interest: metastable $^{88}$Sr. Using atomic parameters
determined in a nearly-{\em ab initio} calculation, we compute adiabatic 
potential energy curves. The anisotropic interatomic interaction, in 
combination with the applied magnetic field, is demonstrated to induce the 
formation of a long-range molecular potential well. This curve correlates to 
two fully polarized, low-field seeking atoms in a rotational s-wave state. 
The coupling among molecular rotational states controls the existence of the 
potential well, and its properties vary as a function of magnetic-field 
strength, thus allowing the scattering length in this state to be tuned. The 
scattering length of metastable $^{88}$Sr displays a resonance at a field of 
$339$~Gauss.
\end{abstract}
\maketitle

\section{Introduction}
\label{sec1}

The ability of van der Waals' equation-of-state of a real gas \cite{HiCu64}
to describe the phase transition from the gaseous to the liquid state is one
of the manifestations of the universal importance that must be attributed to 
interatomic forces. Renewed interest in their detailed understanding has been
prompted by the experimental demonstration \cite{CoWi02,Kett02} of a pure
quantum phase transition---Bose-Einstein condensation---in dilute gases of
alkali-metal atoms. Ultracold conditions, which are necessary for the formation
of a Bose-Einstein condensate, provide an ideal setting to observe, and 
measure precisely, weak interatomic interactions \cite{WeBa99}.

Alkali atoms constitute a natural choice: They possess strong electric dipole 
transitions from their ground state that lie in the visible range of the
electromagnetic spectrum. This makes them ideal candidates for laser cooling
techniques. Furthermore, as effective one-electron systems they have a magnetic
dipole moment and can be trapped in a suitably shaped magnetic field. 
The experimental interest in alkali atoms triggered precision theoretical
studies devoted to uncovering their long-range interaction properties
\cite{MaSa94,MaBa94,PaTa97,DeJo99}.

Noble-gas atoms become accessible to laser cooling if they are not in their
electronic ground state but in an excited, metastable one. Dispersion 
coefficients for metastable helium, $2$~$\phantom{}^{1}S\phantom{}_{0}$ and 
$2$~$\phantom{}^{3}S\phantom{}_{1}$, have been calculated, for example, by 
Chen \cite{Chen95} and by Yan and Babb \cite{YaBa98}. The metastable states 
of the heavier noble gases can be written as $\left(n{\mathrm{p}}^5 
(n+1){\mathrm{s}}\right)$~$\phantom{}^{3}P\phantom{}_{2}$ 
($n = 2$ for neon, and so on). The associated electron distributions are 
nonspherical, in contrast to the $2$~$\phantom{}^{1}S\phantom{}_{0}$ and 
$2$~$\phantom{}^{3}S\phantom{}_{1}$ states of atomic He---and in contrast 
to ground-state alkali atoms. The interactions between such metastable 
noble-gas atoms are not isotropic, and they are not pure dispersion forces. 
(Dispersion coefficients are reported in Ref.~\cite{DeDa00}.) The most 
important new interaction is the electric quadrupole-quadrupole interaction, 
as discussed by Doery {\em et al.} \cite{DoVr98}. Under normal conditions, 
anisotropic interactions between the constituents of a gas tend to be smeared 
out due to thermal averaging. Only ultracold temperatures permit their 
detailed study.

Recently, alkaline-earth atoms have moved into the focus of interest.
In their ground state they can be laser-cooled using the strong 
$\phantom{}^{1}S\phantom{}_{0} \rightarrow \phantom{}^{1}P\phantom{}_{1}$
transition \cite{DiVo99,XuLo02}. Katori {\em et al.} exploited the 
extremely narrow $\phantom{}^{1}S\phantom{}_{0} \rightarrow 
\phantom{}^{3}P\phantom{}_{1}$ intercombination transition to achieve laser 
cooling of ground-state $^{88}$Sr down to a few hundred nano-Kelvin 
\cite{KaId99}. Stable alkaline-earth isotopes with an even number of nucleons 
predominate, which therefore possess no nuclear magnetic moment: Their 
electronic spectra are free of hyperfine structure. This simplifies the 
interpretation of experimental data and facilitates comparison to theory 
\cite{MaJu01}. The analysis of the photoassociative spectrum of cold 
$^{40}$Ca atoms measured by Zinner {\em et al.} may serve as an illustration 
\cite{ZiBi00}.

Ground-state alkaline-earth atoms, which have a closed-shell electronic 
structure, do not lend themselves to pure magnetic trapping. Switching
off the light field employed to hold the laser-cooled atoms in a 
magneto-optical trap, followed by evaporative cooling down to quantum 
degeneracy, is thus not an option. An interesting strategy to 
overcome this obstacle is to work with metastable alkaline-earth atoms in 
their lowest $\left(n{\mathrm{s}}
n{\mathrm{p}}\right)$~$\phantom{}^{3}P\phantom{}_{2}$ state 
\cite{LoBo01,LoBo02}, which have radiative lifetimes of the order of ten
minutes \cite{Dere01}. The experimental feasibility of magnetic trapping of 
metastable $^{88}$Sr has already been demonstrated \cite{KaId01,NaSi03,XuLo03}.

Metastable alkaline-earth atoms share with metastable noble-gas atoms
(excluding He) the property of interacting through anisotropic forces.
However, as was shown quantitatively by Derevianko \cite{Dere01}, the dominant
electric quadrupole-quadrupole interaction is noticeably stronger among
alkaline-earth atoms: Their large quadrupole moment is a consequence of the 
relatively diffuse spatial distribution of the excited p-electron; the 
lack of spherical symmetry of a metastable noble-gas atom, on the other hand,
is due to the hole in the compact valence shell, shielded by an orbiting 
electron of s-symmetry. Since metastable alkaline-earth atoms---in contrast
to noble-gas atoms---do not suffer from trap losses via the phenomenon
of Penning ionization \cite{Sisk93}, they offer the exciting prospect of
quantum degenerate gases with pronounced anisotropic interatomic interactions.

Derevianko and co-workers recently presented a first, simplified analysis of 
the ultracold collision properties of metastable alkaline-earth atoms
\cite{DePo03}. They predict the existence of a long-range molecular 
potential well for the electronic state of highest Zeeman energy in a given
external magnetic field. The scattering length on this potential energy curve
can be tuned by adjusting the magnetic-field strength, thus enabling control
of sign and strength of the effective interatomic interaction at ultracold
temperatures. The numerical results in Ref.~\cite{DePo03} are based on the 
assumption that s-wave scattering of two metastable alkaline-earth atoms in the
low-field seeking state mentioned above is sensitive only to the rotationally
invariant part of the interatomic interaction potential. However, as we shall 
show in this paper, that approximation is not justified in the case of 
relatively strong anisotropic coupling and does not lead to quantitatively 
accurate predictions.

In order to provide a systematic framework for describing anisotropic
interactions, we present in Sec.~\ref{sec2} a detailed tensorial analysis
of the leading interatomic interaction operators. In particular, we show that
the electric quadrupole-quadrupole operator transforms like a spherical 
tensor of rank four, and we decompose the electric dipole-dipole dispersion
operator into tensors of rank zero, two, and four. Similar tensorial ideas
have been applied, for instance, to the anisotropic interactions between 
molecules \cite{WoMu77,Leav80,Piec84} and these were crucial to formally
demonstrate the surprising existence of a vector interaction---a tensorial 
coupling of rank one---between a high-angular momentum Rydberg electron
and an anisotropic ionic core \cite{Zyge90,ClGr99}. General introductions to 
the techniques of spherical tensor algebra may be found in Refs.~\cite{Edmo57},
\cite{FaRa59}, and \cite{Zare88}.

An important new ingredient, facilitated by our use of tensorial methods,
is the inclusion of the quantum-mechanical rotation of an anisotropically 
interacting diatomic system in a magnetic field (Sec.~\ref{sec3}). At 
interatomic distances of a few hundred Bohr radii, electronic interaction 
energies and rotational energies are comparable. Anisotropic interatomic
interaction leads to coupling between different rotational quantum states,
and the classification of a diatomic eigenstate using just a single rotational
quantum number ceases to be meaningful. 

Section~\ref{sec4} is devoted to the discussion of a concrete example: 
metastable $^{88}$Sr. We have calculated all relevant long-range parameters 
(magnetic dipole moment, electric quadrupole moment, and dispersion 
coefficients), treating the correlation between the two valence electrons in 
atomic $^{88}$Sr fully. Using these parameters, which are in generally good
agreement with the ones presented in Ref.~\cite{DePo03}, we obtain adiabatic
potential curves by diagonalizing the complete diatomic Hamiltonian as a 
function of interatomic separation. Our calculations reproduce the phenomenon
of a potential well in the low-field seeking state, correlating at large 
interatomic separations to a rotational s-wave state and fully polarized atoms
with highest Zeeman energy. We make use of perturbation theory to shed light
on how the interplay of the different physical operators generates the 
long-range molecular potential well. Our analysis clarifies how the 
magnetic-field strength affects depth and location of the potential minimum. 
It also shows that, at interatomic distances in the vicinity of and below the 
potential well minimum, the eigenstate acquires an appreciable admixture of 
higher rotational quantum states. Indeed, for this reason the scattering 
lengths we find on the basis of our numerical data differ substantially from 
those reported by Derevianko {\em et al.} \cite{DePo03}.

A concluding discussion is given in Sec.~\ref{sec5}. We use atomic units 
throughout, unless otherwise noted.

\section{Tensorial structure of interatomic interaction operators}
\label{sec2}

\subsection{Leading expansion terms and Feshbach formalism}
\label{sc2sb1}

The interaction Hamiltonian, $H_{\mathrm{int}}$, of two atoms
whose electron clouds do not overlap can be expanded in terms of inverse
powers of the interatomic distance, $R$, by applying techniques well known 
from classical electrodynamics \cite{Jack98,Marg39,DaDa66,Chan67}. All 
operators in the resulting series can be expressed using purely atomic 
observables.

Given a neutral atomic species with nonvanishing magnetic-dipole and  
electric-quadrupole moments, the leading 
expansion contributions up to fifth order in $1/R$ can be written as 
\begin{equation}
\label{s2s1e1}
H_{\mathrm{int}} = H_{\mathrm{dd}} + H_{\mathrm{mm}} + H_{\mathrm{dq}}
+ H_{\mathrm{qq}},
\end{equation}
where 
\begin{equation}
\label{s2s1e2}
H_{\mathrm{dd}} = \frac{1}{R^3}\sum_{i_1,i_2}
\left\{
{\bm x}_{i_1}\cdot{\bm x}_{i_2} - 3 ({\bm x}_{i_1}\cdot{\bm n})
({\bm x}_{i_2}\cdot{\bm n})
\right\}
\end{equation}
is the familiar electric dipole-dipole interaction operator.
${\bm x}_{i_1}$ symbolizes the position of an electron belonging to atom $1$
relative to the nucleus of that atom. Similarly, ${\bm x}_{i_2}$ refers to 
an electron in atom $2$, measured in relation to the nucleus of atom $2$. 
${\bm n}$ denotes a unit vector along the diatomic axis. Using spherical 
multipole moment operators, defined as 
\begin{equation}
\label{s2s1e3}
q_{l,m}^{(k)} := - \sum_{i_k}r_{i_k}^l 
C_{l,m}(\vartheta_{i_k},\varphi_{i_k}),
\end{equation}
$(r_{i_k},\vartheta_{i_k},\varphi_{i_k})$ representing spherical coordinates
of vector ${\bm x}_{i_k}$ and $C_{l,m}(\vartheta_{i_k},\varphi_{i_k})$
being related to the spherical harmonic 
$Y_{l,m}(\vartheta_{i_k},\varphi_{i_k})$ through the simple relation 
\begin{equation}
\label{s2s1e4}
C_{l,m}(\vartheta_{i_k},\varphi_{i_k}) := 
\sqrt{\frac{4 \pi}{2 l + 1}}Y_{l,m}(\vartheta_{i_k},\varphi_{i_k}),
\end{equation}
the electric dipole-dipole operator reads 
\begin{equation}
\label{s2s1e5}
H_{\mathrm{dd}} = - \frac{1}{R^3}\left\{
q_{1,+1}^{(1)} q_{1,-1}^{(2)} + 2 q_{1,0}^{(1)} q_{1,0}^{(2)} 
+ q_{1,-1}^{(1)} q_{1,+1}^{(2)} 
\right\}.
\end{equation}
The basic assumption underlying this representation of $H_{\mathrm{dd}}$
is that the vector ${\bm n}$ introduced in Eq.~(\ref{s2s1e2}) is 
identical to ${\bm e}_z$, the Cartesian unit vector along the $z$-axis
of a chosen reference frame. Hence, Eq.~(\ref{s2s1e5}), and all other
equations in the remainder of this section, refer to a body-fixed frame. 
We will return to this point later when we incorporate the effect of 
molecular rotation.
 
The second term on the right-hand side of Eq.~(\ref{s2s1e1}) describes
the magnetic dipole-dipole interaction \cite{Meat66} between atom $1$ and 
atom $2$:
\begin{equation}
\label{s2s1e6}
H_{\mathrm{mm}} = - \frac{1}{R^3}\left\{
\mu_{+1}^{(1)} \mu_{-1}^{(2)} + 2 \mu_{0}^{(1)} \mu_{0}^{(2)}
+ \mu_{-1}^{(1)} \mu_{+1}^{(2)}
\right\}.
\end{equation}
The magnetic dipole operator, 
\begin{equation}
\label{s2s1e7}
{\bm \mu}^{(k)} = - \mu_{\mathrm{B}} \left\{
{\bm j}^{(k)} + {\bm s}^{(k)}
\right\}
\end{equation}
($\mu_{\mathrm{B}}$: Bohr magneton; ${\bm j}^{(k)}$: total angular momentum
of atom $k$; ${\bm s}^{(k)}$: spin of atom $k$), is a spherical tensor of
rank one. Since we restrict our treatment of magnetic effects to the 
dipole term, we suppress the rank index of ${\bm \mu}^{(k)}$. In fact, at 
interatomic distances of only a few hundred Bohr radii or less, even magnetic 
dipole-dipole coupling is negligible. We include it here because it becomes
the dominant interaction beyond a distance of a thousand atomic units, as we
will show in Sec.~\ref{sec4}.

The expansion term in fourth order contributing to $H_{\mathrm{int}}$ is the
electric dipole-quadrupole operator,
\begin{eqnarray}
\label{s2s1e8}
H_{\mathrm{dq}} & = & - \frac{\sqrt{3}}{R^4}
\left\{q_{1,-1}^{(1)} q_{2,+1}^{(2)} + \sqrt{3} q_{1,0}^{(1)} q_{2,0}^{(2)}
+ q_{1,+1}^{(1)} q_{2,-1}^{(2)}\right. \nonumber \\
& & \left. - q_{2,-1}^{(1)} q_{1,+1}^{(2)} - 
\sqrt{3} q_{2,0}^{(1)} q_{1,0}^{(2)}
- q_{2,+1}^{(1)} q_{1,-1}^{(2)}\right\},
\end{eqnarray}
and finally, in $1/R^5$, we have electric quadrupole-quadrupole coupling:
\begin{eqnarray}
\label{s2s1e9}
H_{\mathrm{qq}} & = & \frac{1}{R^5}
\left\{q_{2,-2}^{(1)} q_{2,+2}^{(2)} + 4 q_{2,-1}^{(1)} q_{2,+1}^{(2)}
+ 6 q_{2,0}^{(1)} q_{2,0}^{(2)}\right. \nonumber \\
& &  \left. + 4 q_{2,+1}^{(1)} q_{2,-1}^{(2)} + q_{2,+2}^{(1)} q_{2,-2}^{(2)}
\right\}.
\end{eqnarray}
Electric dipole-octupole interactions, which are also proportional to 
$1/R^5$, are neglected.

Our focus in this work is on heavy alkaline-earth atoms, in which relativistic
effects are significant even in the valence shell. We will assume that the
splitting between neighboring atomic energy levels corresponding to the same
fine-structure manifold be larger than the interatomic interaction energy.
Let both atoms be in the same well-defined fine-structure state characterized
by a total atomic angular momentum $j$ and a collective quantum number 
$\xi$ taking into account all other electronic degrees-of-freedom, apart from
the projection quantum number $m$ of the total atomic angular momentum.
Without interatomic coupling, direct products of the form 
\[
\left|j,m_1,\xi\right>^{(1)} \left|j,m_2,\xi\right>^{(2)}
\]
span an energetically degenerate, $(2 j + 1)^2$-dimensional subspace of the
full electronic Hilbert space of the diatomic system. This {\em uncoupled}
basis of the degenerate model space formed the starting point of the 
approach taken by other authors \cite{DeDa00,Dere01,DePo03}. However, in 
order to make most efficient use of the powerful techniques of tensor algebra 
and also to demonstrate and exploit inherent symmetries of the problem, we 
prefer to work in a {\em coupled} representation:
\begin{eqnarray}
\label{s2s1e10}
\left|J,M,\Xi\right> & := & \sum_{m} 
C(j,j,J;m,M-m,M) \\
& & \times \left|j,m,\xi\right>^{(1)} \left|j,M-m,\xi\right>^{(2)}.
\nonumber 
\end{eqnarray}
$C(j,j,J;m,M-m,M)$ is a Clebsch-Gordan coefficient \cite{Rose95}, mediating 
the transformation from the uncoupled basis to the coupled one with the total
electronic angular momentum $J$ running from $0$ to $2 j$ and 
$M = -J,-J+1...,J-1,J$. The capital letter $\Xi$ is employed to collectively
symbolize all other electronic quantum numbers of the noninteracting
diatomic system, including for example $j$.

By representing $H_{\mathrm{int}}$ in the $(2 j + 1)^2$-dimensional model space
and diagonalizing the resulting matrix, approximate interaction energies
can be obtained. Note that 
\begin{equation}
\label{s2s1e11}
\left<J,M,\Xi\right|\left.H_{\mathrm{dd}}\right|
J^{''},M^{''},\left.\Xi\right>  = 0
\end{equation}
and 
\begin{equation}
\label{s2s1e12}
\left<J,M,\Xi\right|\left.H_{\mathrm{dq}}\right|
J^{''},M^{''},\left.\Xi\right>  = 0,
\end{equation}
because all atomic eigenstates comprising the model space have the same 
parity and, therefore, are not coupled by the electric dipole operator.
The two atoms experience no direct electric dipole-dipole and 
no direct electric dipole-quadrupole interactions.

However, another important ingredient is still missing: the electric 
dipole-dipole dispersion interaction, which is a consequence of coupling 
to electronic states outside the model space. This can be included by 
defining complementary projection operators
\begin{equation}
\label{s2s1e13}
{\cal P} := \sum_{J,M} \left|J,M,\Xi\right>\left<J,M,\Xi\right|
\end{equation}
and 
\begin{equation}
\label{s2s1e14}
{\cal Q} := {\bm 1} - {\cal P}.
\end{equation}
Let $H_0$ denote the Hamiltonian of the noninteracting system, such that 
\begin{equation}
\label{s2s1e15}
H_0 \left|J,M,\Xi\right> = E_0(\Xi) \left|J,M,\Xi\right>.
\end{equation}
(The eigenenergies of $H_0$ depend only on atomic quantum numbers
$j$ and $\xi$, not on $J$ and $M$.) According to Feshbach \cite{Fesh58,Fesh62},
any eigenenergy $E$ of the interacting system, defined by the Schr\"{o}dinger
equation
\begin{equation}
\label{s2s1e16}
\left\{H_0 + H_{\mathrm{int}}\right\} \left|\Psi\right> = E \left|\Psi\right>,
\end{equation}
can formally be computed within the model space, provided
${\cal P} \left|\Psi\right> \ne 0$. ${\cal P} \left|\Psi\right>$, an element
of the model space, is an eigenvector of the effective Hamiltonian
\begin{eqnarray}
\label{s2s1e17}
H_{\mathrm{eff}} & := & {\cal P} \left\{H_0 + H_{\mathrm{int}}\right\} 
{\cal P} \\
& & + {\cal P} H_{\mathrm{int}} {\cal Q}
\left[E - {\cal Q} \left\{H_0 + H_{\mathrm{int}}\right\} {\cal Q}\right]^{-1}
{\cal Q} H_{\mathrm{int}} {\cal P};
\nonumber
\end{eqnarray}
the associated eigenvalue is $E$:
\begin{equation}
\label{s2s1e18}
H_{\mathrm{eff}} {\cal P} \left|\Psi\right> = E {\cal P} \left|\Psi\right>.
\end{equation}
By making use of Eqs.~(\ref{s2s1e1}), (\ref{s2s1e11}), (\ref{s2s1e12}), and 
(\ref{s2s1e15}), corrections to $E_0(\Xi)$ due to 
interatomic coupling can be found by diagonalizing 
\begin{equation}
\label{s2s1e19}
H_{\mathrm{I}} := H_{\mathrm{eff}} - E_0(\Xi) {\cal P} 
= {\cal P}\left\{
H_{\mathrm{mm}} + H_{\mathrm{qq}} + H_{\mathrm{dis}}
\right\}{\cal P}.
\end{equation}
$H_{\mathrm{dis}}$ is the electric dipole-dipole dispersion interaction 
operator, which is proportional to $1/R^6$:
\begin{equation}
\label{s2s1e20}
H_{\mathrm{dis}} := H_{\mathrm{dd}} {\cal Q}
\left[E_0(\Xi) - {\cal Q} H_0 {\cal Q}\right]^{-1}
{\cal Q} H_{\mathrm{dd}}.
\end{equation}
Expansion terms of higher order in $1/R$ (as well as magnetic dipole-dipole
dispersion coupling) have been dropped in Eq.~(\ref{s2s1e19}).

\subsection{Coupled representation of interatomic interaction operators}
\label{sc2sb2}

Next we address the transformation properties of $H_{\mathrm{I}}$ under
simultaneous rotations of atom $1$ and atom $2$. Since excited 
alkaline-earth atoms are, in general, nonspherical, it is natural to ask 
how the interatomic interaction reflects this lack of isotropy.
To that end we seek a compact tensorial formulation of $H_{\mathrm{mm}}$, 
$H_{\mathrm{qq}}$, and $H_{\mathrm{dis}}$, by coupling the respective 
atomic tensor operators to diatomic ones. This approach also has the 
advantage that it enables a very systematic and elegant evaluation of
matrix elements of $H_{\mathrm{I}}$ with respect to the coupled basis 
introduced in Eq.~(\ref{s2s1e10}).

We first turn our attention to $H_{\mathrm{mm}}$, Eq.~(\ref{s2s1e6}).
Each direct product $\mu_{m}^{(1)} \mu_{-m}^{(2)}$ of atomic tensor operators
can be expanded as follows:
\begin{equation}
\label{s2s2e1}
\mu_{m}^{(1)} \mu_{-m}^{(2)} = \sum_{K=0}^{2} C(1,1,K;m,-m,0) M_{K,0},
\end{equation}
where
\begin{eqnarray}
\label{s2s2e2}
M_{K,0} & := & \left[{\bm \mu}^{(1)} \otimes {\bm \mu}^{(2)}\right]_{K,0} \\
& = & \sum_{m} C(1,1,K;m,-m,0) \mu_{m}^{(1)} \mu_{-m}^{(2)}
\nonumber
\end{eqnarray}
is the $0$-component of the irreducible tensor product of rank $K$ of 
tensors ${\bm \mu}^{(1)}$ and ${\bm \mu}^{(2)}$. Upon inserting 
Eq.~(\ref{s2s2e1}) in Eq.~(\ref{s2s1e6}), it is easily seen that 
\begin{equation}
\label{s2s2e3}
H_{\mathrm{mm}} = - \frac{\sqrt{6}}{R^3} M_{2,0},
\end{equation}
which demonstrates the magnetic dipole-dipole interaction operator is the
$0$-component of a second-rank tensor. A similar argument can be applied
to show that $H_{\mathrm{qq}}$ (Eq.~(\ref{s2s1e9})) is the $0$-component 
of a fourth-rank tensor:
\begin{eqnarray}
\label{s2s2e4}
H_{\mathrm{qq}} & = & \frac{\sqrt{70}}{R^5} Q_{4,0}, \\
\label{s2s2e5}
Q_{4,0} & := & \left[{\bm q}_{2}^{(1)} \otimes {\bm q}_{2}^{(2)}\right]_{4,0}.
\end{eqnarray}

For the purpose of uncovering the tensorial structure of the dispersion 
interaction operator, $H_{\mathrm{dis}}$ (Eq.~(\ref{s2s1e20})), we draw 
on considerations used by Fano and Macek in their study of the angular
distribution and polarization of the light emitted by atoms excited in 
collision processes \cite{FaMa73}. Greene and Zare exploited those ideas 
to describe the anisotropic emission of photofragments \cite{GrZa82} and
of laser-induced fluorescence \cite{GrZa83}.

In analogy to Eq.~(\ref{s2s2e3}), the electric dipole-dipole interaction
operator in Eq.~(\ref{s2s1e20}) is 
\begin{eqnarray}
\label{s2s2e6}
H_{\mathrm{dd}} & = & -\frac{\sqrt{6}}{R^3} D_{2,0}, \\
\label{s2s2e7}
D_{2,0} & := & \left[{\bm q}_{1}^{(1)} \otimes {\bm q}_{1}^{(2)}\right]_{2,0}.
\end{eqnarray}
Furthermore, the operator 
${\cal Q} \left[E_0(\Xi) - {\cal Q} H_0 {\cal Q}\right]^{-1} {\cal Q}$
is a tensor of rank zero:
\begin{equation}
\label{s2s2e8}
O_{0,0} :=
{\cal Q} \left[E_0(\Xi) - {\cal Q} H_0 {\cal Q}\right]^{-1} {\cal Q}.
\end{equation}
To prove this, we employ a more explicit representation of $O_{0,0}$:
\begin{eqnarray}
\label{s2s2e9}
O_{0,0} & = & \sum_{\Xi^{'}\ne\Xi} \frac{1}{E_0(\Xi)-E_0(\Xi^{'})} \\
& & \times
\sum_{J^{'},M^{'}} \left|J\right.^{'},M^{'},\Xi^{'}\left.\right>
\left<J\right.^{'},M^{'},\Xi^{'}\left.\right|. \nonumber
\end{eqnarray}
Each operator $\sum_{M^{'}} \left|J\right.^{'},M^{'},\Xi^{'}\left.\right>
\left<J\right.^{'},M^{'},\Xi^{'}\left.\right|$ is, according to Racah's 
definition of irreducible tensors \cite{RacI42,RaII42,Raca43}, a scalar, 
i.e. a tensor of rank zero, since it is not difficult to see that 
for $m = +1,0,-1$
\begin{equation}
\label{s2s2e10}
\left[J_m, \sum_{M^{'}} \left|J\right.^{'},M^{'},\Xi^{'}\left.\right>
\left<J\right.^{'},M^{'},\Xi^{'}\left.\right|\right] = 0.
\end{equation}
$J_{+1}$, $J_{0}$, and $J_{-1}$ are the spherical tensor components of the 
total electronic angular momentum operator ${\bm J}$. $O_{0,0}$, being a sum 
of scalars, is therefore also a scalar. The dispersion interaction operator 
can now be written as 
\begin{eqnarray}
\label{s2s2e11}
H_{\mathrm{dis}} & = & \frac{6}{R^6} D_{2,0} O_{0,0} D_{2,0} \nonumber \\
& = & \frac{6}{R^6} \sum_{K=0}^{4} C(2,2,K;0,0,0) I_{K,0} \nonumber \\
& = & \frac{6}{R^6} \left\{\sqrt{\frac{1}{5}} I_{0,0} 
- \sqrt{\frac{2}{7}} I_{2,0} + \sqrt{\frac{18}{35}} I_{4,0}\right\},
\end{eqnarray}
the tensor operators $I_{K,0}$ being given by 
\begin{equation}
\label{s2s2e12}
I_{K,0} := \left[\left[{\bm D}_{2} \otimes {\bm O}_{0}\right]_{2}
\otimes {\bm D}_{2}\right]_{K,0}, \enspace K = 0,2,4.
\end{equation}

In this way, a decomposition of the effective interaction operator
$H_{\mathrm{I}}$ into a scalar, a second-rank tensor, and a fourth-rank
tensor has been achieved:
\begin{eqnarray}
\label{s2s2e13}
H_{\mathrm{I}} & = & 
{\cal P}\left\{T_{0,0} + T_{2,0} + T_{4,0}\right\}{\cal P}, \\
T_{0,0} & := & \frac{6}{R^6} \sqrt{\frac{1}{5}} I_{0,0}, 
\nonumber \\
T_{2,0} & := & - \frac{6}{R^6} \sqrt{\frac{2}{7}} I_{2,0}
- \frac{\sqrt{6}}{R^3} M_{2,0},
\nonumber \\
T_{4,0} & := & \frac{6}{R^6} \sqrt{\frac{18}{35}} I_{4,0}
+ \frac{\sqrt{70}}{R^5} Q_{4,0}.
\nonumber
\end{eqnarray}
(The projection operator ${\cal P}$ is a scalar, in view of 
Eq.~(\ref{s2s2e10}).) The physical significance of this result lies in the 
distinct transformation properties of the three different tensor types under 
rotations generated by ${\bm J}$. These are rotations of the two atoms about 
their respective centers. The scalar term, $T_{0,0}$, is invariant under such 
rotations: It describes a purely isotropic interaction. The anisotropy of the 
long-range interaction between two metastable alkaline-earth atoms is a 
consequence of the presence of the two higher-rank tensors. $T_{2,0}$ 
transforms like the spherical harmonic $Y_{2,0}$ (a ``d-orbital''), 
$T_{4,0}$ like $Y_{4,0}$ (a ``g-orbital''). Of course, rotations about the 
$z$-axis---the interatomic axis---leave all tensorial terms of 
$H_{\mathrm{I}}$ invariant.

\subsection{Reduced matrix elements}
\label{sc2sb3}

We next focus on the representation of $H_{\mathrm{I}}$ with respect to the
coupled basis of the model space, Eq.~(\ref{s2s1e10}). By putting to use
the celebrated Wigner-Eckart theorem \cite{Saku94}, the matrix elements of the 
operator $M_{2,0}$, for instance, can be written as 
\begin{widetext}
\begin{equation}
\label{s2s3e1}
\left<J,M,\Xi\right|\left.M_{2,0}\right|J^{''},M^{''},\left.\Xi\right>  =  
\frac{1}{\sqrt{2 J + 1}} C(J^{''},2,J;M^{''},0,M) 
\left<J,\Xi\right. \parallel M_{2} \parallel
J^{''},\left.\Xi\right>. 
\end{equation}
\end{widetext}
The quantity $\left<J,\Xi\right. \parallel M_{2} \parallel
J^{''},\left.\Xi\right>$ is referred to as a {\em reduced matrix element}. 
It is independent of projection quantum numbers and, for that reason, is not 
specific to the body-fixed frame in which these expressions have been 
evaluated.

Using the definitions 
\begin{equation}
\label{s2s3e2}
[A,B,...,Z] := (2 A + 1)(2 B + 1)...(2 Z + 1)
\end{equation}
and
\begin{equation}
\label{s2s3e3}
\left<j,\xi\right. \parallel\mu^{(1)}\parallel \left.j,\xi\right> = 
\left<j,\xi\right. \parallel\mu^{(2)}\parallel \left.j,\xi\right> =:
\left<j,\xi\right. \parallel\mu\parallel \left.j,\xi\right>, 
\end{equation}
standard Wigner-Racah algebra allows a straightforward evaluation of the
reduced matrix elements of the tensor ${\bm M}_{2}$:
\begin{widetext}
\begin{equation}
\label{s2s3e4}
\left<(jj)J,\Xi\right. \parallel 
\left[{\bm \mu}^{(1)} \otimes {\bm \mu}^{(2)}\right]_{2}
\parallel (jj)J^{''},\left.\Xi\right> =
[J,2,J^{''}]^{1/2} \left\{
\begin{array}{ccc}
j & j & 1 \\
j & j & 1 \\
J & J^{''} & 2 \\
\end{array}
\right\}
\left<j,\xi\right. \parallel\mu\parallel \left.j,\xi\right>^2.
\end{equation}
\end{widetext}
The $9$-j symbol \cite{VaMo88} in this expression satisfies the 
equation 
\begin{equation}
\label{s2s3e5}
\left\{
\begin{array}{ccc}
j & j & 1 \\
j & j & 1 \\
J & J^{''} & 2 \\
\end{array}
\right\}
= (-1)^{J + J^{''}}
\left\{
\begin{array}{ccc}
j & j & 1 \\
j & j & 1 \\
J & J^{''} & 2 \\
\end{array}
\right\}.
\end{equation}
This is a simple consequence of the symmetry properties of $9$-j symbols
under interchange of two of its rows and the fact that both atoms have, within
the model space, the same angular momentum, $j$. (The two atoms would not
have to be of the same species, however.) Hence,
$\left<J,\Xi\right. \parallel M_{2} \parallel
J^{''},\left.\Xi\right> = 0$, if $J + J^{''}$ is odd. In other words,
states of even and odd $J$ are not coupled by ${\bm M}_{2}$. We will show
below that this is true for all operators defining $H_{\mathrm{I}}$
(Eq.~(\ref{s2s2e13})).

The reduced matrix element $\left<j,\xi\right. \parallel\mu\parallel 
\left.j,\xi\right>$ in Eq.~(\ref{s2s3e4}) can be related to the atomic
magnetic dipole moment,
\begin{eqnarray}
\label{s2s3e6}
\left<\mu\right> & := & \left<j,j,\xi|\mu_{0}|j,j,\xi\right> \\
& = & \sqrt{\frac{j}{(j + 1)(2 j + 1)}}
\left<j,\xi\right. \parallel\mu\parallel
\left.j,\xi\right>,
\nonumber
\end{eqnarray}
so that
\begin{widetext}
\begin{equation}
\label{s2s3e7}
\left<J,\Xi\right. \parallel M_{2} \parallel J^{''},\left.\Xi\right> =
\sqrt{5} [J,J^{''}]^{1/2} \left\{
\begin{array}{ccc}
j & j & 1 \\
j & j & 1 \\
J & J^{''} & 2 \\
\end{array}
\right\}
\frac{(j + 1)(2 j + 1)}{j} \left<\mu\right>^2.
\end{equation}
\end{widetext}
Equation (\ref{s2s3e7}) is valid only if $j$ is different from zero.
Otherwise, the reduced matrix element vanishes, because in this case
the first and second row of the $9$-j symbol in Eq.~(\ref{s2s3e4}) do not 
satisfy the triangle condition and the $9$-j symbol therefore vanishes. 

Finding the reduced matrix elements of ${\bm Q}_{4}$ is equally easy. We 
introduce the atomic quadrupole moment as 
\begin{eqnarray}
\label{s2s3e8}
\left<q_{2}\right> & := & 2 \left<j,j,\xi|q_{2,0}|j,j,\xi\right> \\
& = & 2 \sqrt{\frac{j (2 j - 1)}{(j + 1)(2 j + 1)(2 j + 3)}}
\left<j,\xi\right. \parallel q_{2} \parallel
\left.j,\xi\right>
\nonumber
\end{eqnarray}
and obtain for $j \ge 1$
\begin{widetext}
\begin{equation}
\label{s2s3e9}
\left<J,\Xi\right. \parallel Q_{4} \parallel J^{''},\left.\Xi\right> =
\frac{3}{4} [J,J^{''}]^{1/2} \left\{
\begin{array}{ccc}
j & j & 2 \\
j & j & 2 \\
J & J^{''} & 4 \\
\end{array}
\right\}
\frac{(j + 1)(2 j + 1)(2 j + 3)}{j (2 j - 1)} \left<q_{2}\right>^2.
\end{equation}
\end{widetext}
Because of properties of the $9$-j symbol, the reduced matrix element
$\left<J,\Xi\right. \parallel Q_{4} \parallel J^{''},\left.\Xi\right>$
vanishes if $j = 0$ or $1/2$ or if $(-1)^{J + J^{''}}$ is negative.

Evaluating the reduced matrix elements of the dispersion interaction
tensors ${\bm I}_{K}$ ($K = 0,2,4$) is slightly more involved. We apply
the well-known rules for determining reduced matrix elements of an 
irreducible tensor product of noncommuting tensor operators and arrive at
\begin{eqnarray}
\label{s2s3e10}
\left<J,\Xi\right. \parallel I_{K} \parallel J^{''},\left.\Xi\right> & = & 
(-1)^{J + J^{''}} \sqrt{2 K + 1}
\sum_{\Xi^{'} \ne \Xi} \sum_{J^{'}}
\left\{
\begin{array}{ccc}
2 & 2 & K \\
J^{''} & J & J^{'} \\
\end{array}
\right\} \\
& & \times
\frac{\left<J,\Xi\right. \parallel D_{2} \parallel J^{'},\Xi^{'}\left.\right>
\left<\right.J^{'},\Xi^{'} \parallel D_{2} \parallel J^{''},\left.\Xi\right>
}{E_0(\Xi)-E_0(\Xi^{'})}. \nonumber
\end{eqnarray}
The coupling by the interaction tensor of rank $K$ between the two electronic
states of angular momentum $J$ and $J^{''}$, respectively, is controlled by
the $6$-j symbol \cite{VaMo88} in Eq.~(\ref{s2s3e10}). In particular,
$K$, $J$, and $J^{''}$ must satisfy the triangle condition 
$|J - K| \le J^{''} \le J + K$. The reduced matrix elements of ${\bm D}_{2}$ 
require a treatment similar to the one that led to Eq.~(\ref{s2s3e4}). Before 
writing down the final result for 
$\left<J,\Xi\right. \parallel I_{K} \parallel J^{''},\left.\Xi\right>$,
however, we need to discuss the notation we will employ. 

Let $\varepsilon(j^{(k)},\xi^{(k)},j,\xi)$ denote the transition energy 
associated with the transition of atom $k$ from the state 
$\left|\right.j,\xi\left>\right.$
to $\left|\right.j^{(k)},\xi^{(k)}\left>\right.$, and 
$f(j^{(k)},\xi^{(k)},j,\xi)$ be the corresponding dipole oscillator strength 
\cite{YaBa96}:
\begin{eqnarray}
\label{s2s3e11}
f(j^{(k)},\xi^{(k)},j,\xi) & := & \frac{2}{3} 
\frac{(-1)^{j - j^{(k)}}}{2 j + 1}
\varepsilon(j^{(k)},\xi^{(k)},j,\xi) \\
& & \times \left<j,\xi\right. \parallel q_{1} \parallel
j^{(k)},\xi^{(k)}\left.\right>
\left<\right.j^{(k)},\xi^{(k)} \parallel q_{1} \parallel
\left.j,\xi\right>. 
\nonumber
\end{eqnarray}
We use these to define an intermediate dispersion coefficient
\begin{eqnarray}
\label{s2s3e12}
B(j^{(1)},j^{(2)},j,\xi) & := & (-1)^{1 + j^{(1)} - j^{(2)}} \\
& & \times 
\sum_{\xi^{(1)}}\phantom{}^{'} \sum_{\xi^{(2)}}\phantom{}^{'}
\frac{f(j^{(1)},\xi^{(1)},j,\xi) f(j^{(2)},\xi^{(2)},j,\xi)}{
\left\{\varepsilon(j^{(1)},\xi^{(1)},j,\xi) + 
\varepsilon(j^{(2)},\xi^{(2)},j,\xi)
\right\}\varepsilon(j^{(1)},\xi^{(1)},j,\xi)
\varepsilon(j^{(2)},\xi^{(2)},j,\xi)
}.
\nonumber
\end{eqnarray}
$\sum_{\xi^{(k)}}^{'}$ indicates a sum over all atomic eigenstates
with angular momentum $j^{(k)}$, excluding $\xi^{(k)} = \xi$ if $j^{(k)} = j$.
The relation between our $B$-coefficients and the intermediate 
$C_{6}$-coefficients used by Derevianko {\em et al.} \cite{DePo03} is
\begin{equation}
\label{s2s3e13}
C_{6}^{j^{(1)} j^{(2)}} = - \frac{27}{8} (2 j + 1)^2 B(j^{(1)},j^{(2)},j,\xi).
\end{equation}

On physical grounds we expect only those $j^{(k)}$ to play a role that are
consistent with dipole-allowed transitions. This fact is expressed by the 
triangle conditions that must be satisfied in the function
\begin{widetext}
\begin{equation}
\label{s2s3e14}
A(J,J^{''},K,j,j^{(1)},j^{(2)}) := \sum_{J^{'}} (2 J^{'} + 1)
\left\{
\begin{array}{ccc}
2 & 2 & K \\
J^{''} & J & J^{'} \\
\end{array}
\right\}
\left\{
\begin{array}{ccc}
j & j^{(1)} & 1 \\
j & j^{(2)} & 1 \\
J & J^{'} & 2 \\
\end{array}
\right\}
\left\{
\begin{array}{ccc}
j^{(1)} & j & 1 \\
j^{(2)} & j & 1 \\
J^{'} & J^{''} & 2 \\
\end{array}
\right\}.
\end{equation}
\end{widetext}
Specifically, $j$, $j^{(k)}$, and $1$ must form a triangle for $A$ not to 
vanish. 

Using definitions (\ref{s2s3e12}) and (\ref{s2s3e14}), the reduced matrix 
elements of ${\bm I}_{K}$ assume a rather compact form:
\begin{eqnarray}
\label{s2s3e15}
\left<J,\Xi\right. \parallel I_{K} \parallel J^{''},\left.\Xi\right> & = &
\frac{45}{4} (2 j + 1)^2 [J,K,J^{''}]^{1/2} \\
& & \times \sum_{j^{(1)}} \sum_{j^{(2)}}
A(J,J^{''},K,j,j^{(1)},j^{(2)}) B(j^{(1)},j^{(2)},j,\xi).
\nonumber
\end{eqnarray}
It is interesting to see why 
$\left<J,\Xi\right. \parallel I_{K} \parallel J^{''},\left.\Xi\right>$
vanishes if $J + J^{''}$ is odd. First note that 
$B(j^{(1)},j^{(2)},j,\xi)$ is symmetric under interchange of $j^{(1)}$ and
$j^{(2)}$, since $\xi^{(1)}$ and $\xi^{(2)}$ are just dummy summation 
variables, which may be renamed. (This statement relies on the premise
that both atoms are identical and described by the same unperturbed quantum 
state $\left|\right.j,\xi\left>\right.$.) An immediate consequence of this 
observation and the dipole-selection rules is the number of independent
B-coefficients: For $j \ge 1$ there are six, corresponding to the 
combinations $\{j^{(1)} = j+1,j^{(2)} = j+1\}$, 
$\{j^{(1)} = j+1,j^{(2)} = j\}$, $\{j^{(1)} = j+1,j^{(2)} = j-1\}$, 
$\{j^{(1)} = j,j^{(2)} = j\}$, $\{j^{(1)} = j,j^{(2)} = j-1\}$, and 
$\{j^{(1)} = j-1,j^{(2)} = j-1\}$.
Second, by exploiting the symmetry properties of the $9$-j symbols in
Eq.~(\ref{s2s3e14}), it can be shown that 
\begin{equation}
\label{s2s3e16}
A(J,J^{''},K,j,j^{(2)},j^{(1)}) = 
(-1)^{J+J^{''}} A(J,J^{''},K,j,j^{(1)},j^{(2)}).
\end{equation}
Hence,
\begin{eqnarray}
\label{s2s3e17}
& & \sum_{j^{(1)}} \sum_{j^{(2)}}
A(J,J^{''},K,j,j^{(1)},j^{(2)}) B(j^{(1)},j^{(2)},j,\xi) \\
& = & \sum_{j^{(2)}} \sum_{j^{(1)}}
A(J,J^{''},K,j,j^{(2)},j^{(1)}) B(j^{(2)},j^{(1)},j,\xi)
\nonumber \\
& = & (-1)^{J+J^{''}} \sum_{j^{(1)}} \sum_{j^{(2)}}
A(J,J^{''},K,j,j^{(1)},j^{(2)}) B(j^{(1)},j^{(2)},j,\xi).
\nonumber
\end{eqnarray}
$\sum_{j^{(1)}} \sum_{j^{(2)}}
A(J,J^{''},K,j,j^{(1)},j^{(2)}) B(j^{(1)},j^{(2)},j,\xi)$ in 
Eq.~(\ref{s2s3e15}) is therefore identical to zero if $J+J^{''}$ is odd.

\section{Molecular rotations in external magnetic field}
\label{sec3}

\subsection{Invariant formulation of interatomic 
interaction Hamiltonian}
\label{sc3sb1}

Suppose each of the two atoms were in a spherically symmetric electronic state
($j = 0$, $m = 0$), defined with respect to a quantization axis held fixed
in the laboratory frame. In this case, the interatomic interaction energy 
clearly does not depend on the orientation of the diatomic axis relative to the
quantization axis. If, however, the atoms are nonspherical ($j > 0$), then
the anisotropic forces, as expressed through the tensorial structure of the 
effective interaction operator $H_{\mathrm{I}}$ in Eq.~(\ref{s2s2e13}), 
attempt to align the atomic angular momenta along the diatomic axis. The
relative orientation of diatomic axis and laboratory-fixed quantization axis
now matters. This leads to a coupling between electronic degrees-of-freedom 
and molecular rotations. In order to put this intuitive picture into more 
quantitative terms, we need a formulation of $H_{\mathrm{I}}$ that makes use 
only of observables specific to the laboratory frame.

Let $U$ be a rotation operator from the laboratory $z$-axis to the 
body-fixed $z$-axis. This operator connects each tensor $T_{K,0}$
in Eq.~(\ref{s2s2e13}) with the $0$-component of the tensor ${\bm T}_{K}$
measured in the laboratory frame:
\begin{equation}
\label{s3s1e1}
T_{K,0}^{(\mathrm{Lab})} := U^{-1} T_{K,0} U.
\end{equation}
On the other hand, the operator $U T_{K,0}^{(\mathrm{Lab})} U^{-1}$ is
just a linear combination of tensor components $T_{K,M}^{(\mathrm{Lab})}$,
$M=-K,...,+K$. The expansion coefficients are the elements of the 
appropriate rotation matrix:
\begin{equation}
\label{s3s1e2}
T_{K,0} = \sum_{M} {\cal D}_{M,0}^{(K)}(\varphi,\vartheta,0) 
T_{K,M}^{(\mathrm{Lab})}.
\end{equation}
Here, $\vartheta$ and $\varphi$ are the polar and the azimuthal angle, 
respectively, of the diatomic axis in the laboratory frame. Since for
integer $K$
\begin{equation}
\label{s3s1e3}
{\cal D}_{M,0}^{(K)}(\varphi,\vartheta,0) = (-1)^{M} 
C_{K,-M}(\vartheta,\varphi)
\end{equation}
(see, for example, Ref.~\cite{Rose95}), the expansion coefficients in
Eq.~(\ref{s3s1e2}) can also be interpreted as the components of a spherical 
tensor of rank $K$, defined with respect to rotations generated by ${\bm L}$, 
the angular momentum operator of the diatomic axis. Thus,
\begin{equation}
\label{s3s1e4}
T_{K,0} = \sqrt{2 K + 1} \left[{\bm C}_{K} \otimes {\bm T}_{K}\right]_{0,0},
\end{equation}
which demonstrates $T_{K,0}$ is a tensor of rank zero with respect to 
rotations generated by the total angular momentum 
${\bm J}_{\mathrm{tot}} = {\bm L} + {\bm J}$. Under rotations of all 
particles---electrons and nuclei---$H_{\mathrm{I}}$ is therefore totally
invariant:
\begin{equation}
\label{s3s1e5}
H_{\mathrm{I}} = \sum_{K=0,2,4} \sqrt{2 K + 1} 
\left[{\bm C}_{K} \otimes {\cal P}{\bm T}_{K}{\cal P}\right]_{0,0}.
\end{equation}

Let $\left|L,M_{L}\right>$ denote an eigenstate of ${\bm L}^2$ and $L_{z}$.
Eigenstates of ${\bm J}_{\mathrm{tot}}^2$ and $J_{\mathrm{tot},z}$ are then 
generated by coupling nuclear and electronic angular momentum states as 
usual:
\begin{eqnarray}
\label{s3s1e6}
\left|J_{\mathrm{tot}},M_{\mathrm{tot}},L,J\right> & := & \sum_{M_{L}}
C(L,J,J_{\mathrm{tot}};M_{L},M_{\mathrm{tot}}-M_{L},M_{\mathrm{tot}}) \\
& & \times \left|L,M_{L}\right> \left|J,M_{\mathrm{tot}}-M_{L},\Xi\right>.
\nonumber
\end{eqnarray}
We next determine the representation of $H_{\mathrm{I}}$ in this basis.
Application of the Wigner-Eckart theorem gives
\begin{widetext}
\begin{equation}
\label{s3s1e7}
\left<J_{\mathrm{tot}},M_{\mathrm{tot}},L,J\right|H_{\mathrm{I}}
\left|\right.J^{''}_{\mathrm{tot}},M^{''}_{\mathrm{tot}},
L^{''},J^{''}\left.\right> = 
\frac{
\delta_{J_{\mathrm{tot}},J^{''}_{\mathrm{tot}}}
\delta_{M_{\mathrm{tot}},M^{''}_{\mathrm{tot}}}
}{\sqrt{2 J_{\mathrm{tot}} + 1}} 
\left<J_{\mathrm{tot}},L,J\right. \parallel H_{\mathrm{I}}
\parallel J_{\mathrm{tot}},L^{''},J^{''}\left.\right>.
\end{equation}
\end{widetext}
Of course, for the scalar tensor $H_{\mathrm{I}}$, $J_{\mathrm{tot}}$ and 
$M_{\mathrm{tot}}$ are good quantum numbers. 

$L$ and $J$, however, are typically not conserved. This is most easily seen
by inspection of the reduced matrix element in Eq.~(\ref{s3s1e7}): 
\begin{eqnarray}
\label{s3s1e8}
\left<J_{\mathrm{tot}},L,J\right. \parallel H_{\mathrm{I}}
\parallel J_{\mathrm{tot}},L^{''},J^{''}\left.\right>
& = & (-1)^{J_{\mathrm{tot}}+J+L^{''}} \sqrt{2 J_{\mathrm{tot}} + 1} \\
& & \times \sum_{K=0,2,4} 
\left\{
\begin{array}{ccc}
L & L^{''} & K \\
J^{''} & J & J_{\mathrm{tot}} \\
\end{array}
\right\}
\nonumber \\
& & \times 
\left< L \right. \parallel C_{K} \parallel L^{''} \left.\right>
\left<J,\Xi\right. \parallel T_{K} \parallel J^{''},\left.\Xi\right>.
\nonumber
\end{eqnarray}
The reduced matrix element of ${\bm C}_{K}$ is that of a renormalized 
spherical harmonic (cf. Eq.~(\ref{s2s1e4})) and is well known \cite{VaMo88}:
\begin{equation}
\label{s3s1e9}
\left< L \right. \parallel C_{K} \parallel L^{''} \left.\right> 
= \sqrt{2 L^{''} + 1} C(L^{''},K,L;0,0,0).
\end{equation}
Hence, it vanishes if $L+L^{''}$ is odd. Otherwise, for $K = 2$ and $4$, 
there can be coupling between rotational states of different quantum numbers 
$L$ and $L^{''}$, provided $L$, $L^{''}$, and $K$ form a triangle. This
coupling exists only if the atomic angular momentum $j$ is larger than zero,
because only then do 
$\left<J,\Xi\right. \parallel T_{2} \parallel J^{''},\left.\Xi\right>$
($j \ge 1/2$)
and 
$\left<J,\Xi\right. \parallel T_{4} \parallel J^{''},\left.\Xi\right>$
($j \ge 1$) in general not vanish. The reduced matrix elements of 
${\bm T}_{K}$ for $K=0,2,4$ follow immediately by combining 
Eqs.~(\ref{s2s2e13}), (\ref{s2s3e7}), (\ref{s2s3e9}), and (\ref{s2s3e15}),
and from our analysis in Sec.\ref{sc2sb3} we already know that $H_{\mathrm{I}}$
conserves $(-1)^J$, but not the total electronic angular momentum $J$ itself.

\subsection{Zeeman operator and rotation energy}
\label{sc3sb2}

Precise experimental studies of long-range interactions can be carried out
with cold, magnetic atoms trapped by means of suitable magnetic field 
configurations \cite{BeEr87}. The presence of an externally applied magnetic 
field ${\bm B}$ breaks the local rotational invariance, and $J_{\mathrm{tot}}$
is no longer expected to be a good quantum number. To model this situation, we
make the simplifying assumption that ${\bm B}$ is homogeneous: 
${\bm B} = {\cal B}{\bm e}_z$. The complete Hamiltonian describing the relative
motion of two interacting atoms, with reduced mass $m_{\mathrm{red}}$, exposed
to a magnetic field is thus
\begin{widetext}
\begin{equation}
\label{s3s2e1}
H = -\frac{1}{2 m_{\mathrm{red}}} \left\{\frac{\partial^{2}}{\partial R^{2}}
+ \frac{2}{R}\frac{\partial}{\partial R}\right\}
+ \frac{{\bm L}^2}{2 m_{\mathrm{red}} R^{2}} + H_{\mathrm{I}}
-({\bm \mu}^{(1)} + {\bm \mu}^{(2)})_{0} {\cal B}.
\end{equation}
\end{widetext}
The internal energy, $E_0(\Xi)$, of the two atoms (see Eq.~(\ref{s2s1e15}))
has been set to zero.

By representing $H$ in the coupled basis defined in Eq.~(\ref{s3s1e6}),
a natural diabatic representation is obtained that can be used for rigorous
scattering calculations. Alternatively, the sum of the matrices of the 
rotational kinetic energy operator, 
$\frac{{\bm L}^2}{2 m_{\mathrm{red}} R^{2}}$, the interatomic
interaction operator, $H_{\mathrm{I}}$, and the Zeeman operator,
$-({\bm \mu}^{(1)} + {\bm \mu}^{(2)})_{0} {\cal B}$ can be diagonalized as 
a function of $R$. This approach yields adiabatic potential energy curves.
The dynamics on these curves is driven by the radial kinetic energy operator,
$-\frac{1}{2 m_{\mathrm{red}}} \left\{\frac{\partial^{2}}{\partial R^{2}}
+ \frac{2}{R}\frac{\partial}{\partial R}\right\}$.

The calculation of the matrix elements of 
$\frac{{\bm L}^2}{2 m_{\mathrm{red}} R^{2}}$ is a trivial matter,
\begin{equation}
\label{s3s2e2}
\left<J_{\mathrm{tot}},M_{\mathrm{tot}},L,J\right|
\frac{{\bm L}^2}{2 m_{\mathrm{red}} R^{2}}
\left|\right.J^{''}_{\mathrm{tot}},M^{''}_{\mathrm{tot}},
L^{''},J^{''}\left.\right> =
\frac{L(L + 1)}{2 m_{\mathrm{red}} R^{2}}
\delta_{J_{\mathrm{tot}},J^{''}_{\mathrm{tot}}}
\delta_{M_{\mathrm{tot}},M^{''}_{\mathrm{tot}}}
\delta_{L,L^{''}} \delta_{J,J^{''}},
\end{equation}
and the treatment of $H_{\mathrm{I}}$ was the subject of the previous 
subsection. The Zeeman operator is discussed in the following. Again,
we utilize Wigner-Racah algebra by noting that 
$-({\bm \mu}^{(1)} + {\bm \mu}^{(2)})_{0} {\cal B}$ is the $0$-component
of a rank-one tensor. It is independent of the orientation of the diatomic
axis and therefore does not couple rotational states of different $L$.

Interestingly, at least in the case of identical atoms, it also does not couple
electronic states of different $J$. To understand this, consider the 
electronic reduced matrix elements of ${\bm \mu}^{(1)} + {\bm \mu}^{(2)}$
($j > 0$):
\begin{eqnarray}
\label{s3s2e3}
\left<J,\Xi\right. \parallel \mu^{(1)} + \mu^{(2)} \parallel
J^{''},\left.\Xi\right> & = & 
(-1)^{J + 2 j + 1} \left((-1)^{J+J^{''}} + 1\right) [J,J^{''}]^{1/2} \\
& & \times
\left\{
\begin{array}{ccc}
j & j & 1 \\
J^{''} & J & j \\
\end{array}
\right\}
\sqrt{\frac{(j + 1)(2 j + 1)}{j}} \left<\mu\right>. \nonumber
\end{eqnarray}
Owing to the factor $(-1)^{J+J^{''}} + 1$, the reduced matrix element vanishes
if $J + J^{''}$ is odd. In addition, however, the $6$-j symbol in 
Eq.~(\ref{s3s2e3}) differs from zero only if $|J-1| \le J^{''} \le J + 1$. 
Since, for atoms of the same species, $J$ and $J^{''}$ are integers, it can be
concluded that $J$ must equal $J^{''}$. The physical origin of this selection
rule is the tensorial rank of the magnetic dipole operator.

Using this result, it is now straightforward to write down the matrix elements
of the Zeeman operator:
\begin{eqnarray}
\label{s3s2e4}
\left<J_{\mathrm{tot}},M_{\mathrm{tot}},L,J\right|
-({\bm \mu}^{(1)} + {\bm \mu}^{(2)})_{0} {\cal B}
\left|\right.J^{''}_{\mathrm{tot}},M^{''}_{\mathrm{tot}},
L^{''},J^{''}\left.\right>  =  (-1)^{J_{\mathrm{tot}} + L + J}
\delta_{M_{\mathrm{tot}},M^{''}_{\mathrm{tot}}}
\delta_{L,L^{''}} \delta_{J,J^{''}} \\
\times 
C(J^{''}_{\mathrm{tot}},1,J_{\mathrm{tot}};M_{\mathrm{tot}},0,M_{\mathrm{tot}})
\sqrt{2 J^{''}_{\mathrm{tot}} + 1}
\left\{
\begin{array}{ccc}
J & J & 1 \\
J^{''}_{\mathrm{tot}} & J_{\mathrm{tot}} & L \\
\end{array}
\right\}
\left<J,\Xi\right. \parallel \mu^{(1)} + \mu^{(2)} \parallel
\left.J,\Xi\right> {\cal B}.
\nonumber
\end{eqnarray}
Thus we have shown that the matrix representation of the Zeeman operator
is diagonal in all angular momentum quantum numbers except $J_{\mathrm{tot}}$.
The Clebsch-Gordan coefficient (as well as the $6$-j symbol) in 
Eq.~(\ref{s3s2e4}) imposes a restriction on the total angular momentum 
quantum numbers that can be coupled: $J_{\mathrm{tot}}$ and 
$J^{''}_{\mathrm{tot}}$ may not differ by more than one unit.

\section{Application to metastable strontium}
\label{sec4}

\subsection{Atomic parameters}
\label{sc4sb1}

In this section we present an application of the theory developed in 
Secs.~\ref{sec2} and \ref{sec3} to collisions between cold, metastable 
$^{88}$Sr atoms. The mass of $^{88}$Sr \cite{AuWa93} is 
$1.60280 \times 10^{5}$~a.u. The other atom-specific parameters
needed are the magnetic dipole moment, $\left<\mu\right>$ (Eq.~(\ref{s2s3e6})),
the electric quadrupole moment, $\left<q_{2}\right>$ (Eq.~(\ref{s2s3e8})),
and the intermediate dispersion coefficients 
$B(j^{(1)},j^{(2)},j,\xi) =: B_{j^{(1)},j^{(2)}}$ (Eq.~(\ref{s2s3e12})).
Here, $j = 2$, and $\xi$ is specified in terms of the other quantum numbers
characterizing the 
$\left(5{\mathrm{s}}5{\mathrm{p}}\right)$~$\phantom{}^{3}P\phantom{}_{2}$
state of atomic strontium. We have performed semiempirical electronic-structure
calculations to determine $\left<\mu\right>$, $\left<q_{2}\right>$, and 
$B_{j^{(1)},j^{(2)}}$.

\begin{table}
\caption[]{Parameters needed for a quantitative characterization of the
long-range interaction between two metastable
$\left(5{\mathrm{s}}5{\mathrm{p}}\right)$~$\phantom{}^{3}P\phantom{}_{2}$
strontium atoms. $\left<\mu\right>$ (Eq.~(\ref{s2s3e6})) is the magnetic
dipole moment, $\left<q_{2}\right>$ (Eq.~(\ref{s2s3e8})) the electric
quadrupole moment, and the coefficients $B_{j^{(1)},j^{(2)}}$
(Eq.~(\ref{s2s3e12})) are needed to describe electric dipole-dipole
dispersion coupling. The subscripts indicate that in the dispersion interaction
process one atom makes a virtual dipole transition from $j=2$ to $j^{(1)}$
and the other one from $j=2$ to $j^{(2)}$. The parameters obtained in this
work were calculated within a semiempirical approach. Derevianko {\em et al.}
\cite{DePo03} presented {\em ab initio} results; their intermediate dispersion
coefficients have been converted using Eq.~(\ref{s2s3e13}). The one-standard
deviation uncertainties cited in Ref.~\cite{DePo03} are given in parentheses.
All data are in atomic units.
}
\label{tab1}
\begin{tabular}{c|c|c}
 & This work & Ref.~\cite{DePo03} \\
\hline
$\left<\mu\right>$ & -3.00 & \\
$\left<q_{2}\right>$ & 15.4 & 15.6(5) \\
$B_{1,1}$ & -132 & -158(16) \\
$B_{2,1}$ & 187 & 203(20) \\
$B_{2,2}$ & -266 & -264(26) \\
$B_{3,1}$ & -343 & -415(42) \\
$B_{3,2}$ & 497 & 555(56) \\
$B_{3,3}$ & -1020 & -1290(130) \\
\end{tabular}
\end{table}

Atomic strontium can be regarded as a two-electron system, at least on the 
excitation energy scale of just a few electron volts relevant to the present
study. This realization can be exploited to efficiently compute and 
successfully reproduce the amazing complexity of photoabsorption Rydberg 
spectra of alkaline-earth atoms. A detailed and general description of the 
underlying concepts, formal developments, and applications is given in 
Ref.~\cite{AyGr96}.

In a first step, we concentrate on the one-electron physics of Sr$^{+}$.
The single valence electron moves in the field of a closed-shell ionic core.
The effective, spherically symmetric potential the electron experiences is
assumed to be of the form
\begin{widetext}
\begin{equation}
\label{s4s1e1}
{\cal V}_{l}(r) = -\frac{1}{r}\left\{2 + (Z-2)\exp{(-\alpha_{l,1}r)}
+ \alpha_{l,2} r \exp{(-\alpha_{l,3}r)}\right\}
- \frac{\alpha_{\mathrm{cp}}}{2 r^{4}}
\left\{1 - \exp{[-(r/r_{l})^{6}]}\right\}.
\end{equation}
\end{widetext}
This one-electron potential is physically well motivated: At large distances
from the ionic core, the electron feels the attraction by a practically 
pointlike Sr$^{++}$ ion. As the electron comes closer, the ionic core responds
to the presence of the electron and becomes polarized. 
$\alpha_{\mathrm{cp}} = 7.5$~a.u. is the dipole polarizability of Sr$^{++}$
\cite{JoKo83}. Below the empirical cutoff radius $r_{l}$, there
is a transition, mediated by the term proportional to $\alpha_{l,2}$,
from the exterior to the interior region of the ionic core. The electron
interacts with an unscreened nucleus of charge $Z = 38$ at $r$ much smaller
than $1/\alpha_{l,1}$. Note that the parameters $\alpha_{l,i}$ and
$r_{l}$ are assumed to be dependent on the orbital angular
momentum quantum number $l$ of the valence electron. They are determined by
comparing the theoretical excitation spetrum computed on the basis of 
Eq.~(\ref{s4s1e1}) with experimental data on Sr$^{+}$; the values used
in this work were taken from Ref.~\cite{AyGr96}.

Another important interaction operator must be included for a quantitative
description of heavy alkaline-earth atoms: spin-orbit interaction, which we
use in the form \cite{CoSh67}
\begin{equation}
\label{s4s1e2}
{\cal V}_{l}^{(\mathrm{so})} = \frac{{\bm s}\cdot{\bm l}}{2 c^2} \frac{1}{r}
\frac{\mathrm{d}{\cal V}_{l}}{\mathrm{d}r}
\left[1-\frac{{\cal V}_{l}}{2 c^2}\right]^{-2}.
\end{equation}
The factor in brackets counteracts the $r^{-3}$-divergence of 
$r^{-1} \mathrm{d}{\cal V}_{l} / \mathrm{d}r$ near the nucleus. ${\bm s}$ 
in Eq.~(\ref{s4s1e2}) is the spin of the valence electron, and $c$ is the 
speed of light. ${\cal V}_{l}^{(\mathrm{so})}$ was adapted by 
Ref.~\cite{Gree90} into the framework of eigenchannel R-matrix calculations 
and utilized to compute precise photodetachment spectra of the heavy 
alkali-metal anions.

The quantum-mechanical motion of an electron exposed to the potential 
${\cal V}_{l}+{\cal V}_{l}^{(\mathrm{so})}$ is solved using a finite-element 
basis \cite{BrSc93,AcSh96,MeGr97} for the radial degree-of-freedom.
A radial box size of $25$ Bohr radii and $600$ finite-element functions,
corresponding to a quadratically spaced grid of $200$ sectors inside the 
box and three functions in each sector, have proved to be sufficient.
The Sr$^{+}$ model Hamiltonian $h$, which comprises kinetic energy and 
effective potential, is represented and diagonalized in a product basis  
of finite-element and spin-angular \cite{Saku94} functions up to $l = 6$
(i-orbitals). For each combination of orbital and total angular momentum
quantum numbers of the single valence electron, the $18$ energetically 
lowest radial eigenfunctions---excluding of course the filled inner-shell 
states---are selected. A two-electron basis set is then constructed by 
coupling pairs of the selected eigenstates of the one-electron Hamiltonian $h$,
performing the angular momentum algebra within the $jj$-coupling scheme.
Denoting the electron-electron Coulomb repulsion as $1/r_{12}$, the
eigenenergies and eigenstates of the valence shell of atomic strontium are 
obtained by diagonalizing the matrix representation of 
$h(1) + h(2) + 1/r_{12}$ in the two-electron basis set. In this way the 
complicated dynamics of the two correlated valence electrons is treated to a 
high degree of accuracy. 

\begin{table}
\caption[]{Order-of-magnitude estimates of the energies, in atomic units,
associated with magnetic dipole-dipole coupling ($E_{\mathrm{mm}}$),
electric quadrupole-quadrupole coupling ($E_{\mathrm{qq}}$), electric
dipole-dipole dispersion interaction ($E_{\mathrm{dis}}$), and molecular
rotation ($E_{\mathrm{rot}}$) of two metastable strontium atoms separated
by a distance of $R$ Bohr radii.}
\label{tab2}
\begin{tabular}{c|c|c|c|c}
$R$ & $E_{\mathrm{mm}}$ & $E_{\mathrm{qq}}$ & $E_{\mathrm{dis}}$ &
$E_{\mathrm{rot}}$ \\
\hline
$10$ & $10^{-7}$ & $10^{-3}$ & $10^{-3}$ & $10^{-7}$ \\
$10^{2}$ & $10^{-10}$ & $10^{-8}$ & $10^{-9}$ & $10^{-9}$ \\
$10^{3}$ & $10^{-13}$ & $10^{-13}$ & $10^{-15}$ & $10^{-11}$ \\
$10^{4}$ & $10^{-16}$ & $10^{-18}$ & $10^{-21}$ & $10^{-13}$ \\
\end{tabular}
\end{table}

The direct numerical evaluation of Eqs.~(\ref{s2s3e6}), (\ref{s2s3e8}), 
and (\ref{s2s3e12}) using the two-electron eigenvectors is straightforward.
Our results for the atomic parameters, which are converged with respect
to all basis set parameters described above, are shown in Table~\ref{tab1},
together with the {\em ab initio} data presented by Derevianko {\em et al.}
\cite{DePo03}. The magnetic dipole moment of metastable $^{88}$Sr was not
explicitly discussed in Ref.~\cite{DePo03}. In a 
$\phantom{}^{3}P\phantom{}_{2}$ state it seems natural to assume an atomic
spin of $1$, consistent with a magnetic dipole moment of $-3$. However,
due to spin-orbit interaction the total spin is not really a good quantum 
number, and it had to be checked that the degree of spin symmetry violation
is negligible. This is, in fact, the case. Our calculated electric quadrupole 
moment agrees well with that of Ref.~\cite{DePo03}; the intermediate 
dispersion coefficients also agree, to within the $10\%$ one-standard 
deviation uncertainty of the data quoted by Derevianko and co-workers 
\cite{DePo03}.

\subsection{Diatomic potential energy curves and scattering lengths}
\label{sc4sb2}

Before we turn our attention to quantitative results on the interaction 
energies between metastable Sr atoms, it is helpful to explore the relative
importance of the different interaction operators we have analyzed in 
Secs.~\ref{sec2} and \ref{sec3}. To that end we have collected in 
Table~\ref{tab2} simple order-of-magnitude estimates as a function of 
interatomic separation. They are easily obtained by using the parameters
in Table~\ref{tab1} and diagonalizing the individual terms the interatomic 
interaction operator, $H_{\mathrm{I}}$ (Eq.~(\ref{s2s2e13})), consists of,
represented in the coupled basis of the electronic model space, 
Eq.~(\ref{s2s1e10}). 

Table~\ref{tab2} allows us to draw several important conclusions.
Between distances of $100$ and $1000$ Bohr radii the electric 
quadrupole-quadrupole interaction dominates. Only as $R=10$ is approached,
the electric dipole-dipole dispersion interaction becomes comparable.
However, at these relatively small distances the interaction energies are
comparable with the fine-structure splitting between the 
$\left(5{\mathrm{s}}5{\mathrm{p}}\right)$~$\phantom{}^{3}P\phantom{}_{2}$
and the 
$\left(5{\mathrm{s}}5{\mathrm{p}}\right)$~$\phantom{}^{3}P\phantom{}_{1}$
state, which is of order $10^{-3}$ Hartree. In other words, at distances
much lower than $100$ Bohr radii the electronic model space we have chosen
in Sec.~\ref{sec2} is no longer appropriate. Truncating all interaction
operators of orders higher than six in $1/R$ (Eq.~(\ref{s2s1e19})) also 
becomes highly questionable under these circumstances. At $R \ge 100$~a.u.,
however, the formalism developed in this paper may be expected to be useful 
for quantitative predictions. 

In the long-range limit, at $R$ larger than $1000$ Bohr radii, magnetic 
dipole-dipole coupling remains the only relevant interatomic interaction
mechanism. Also shown in Table~\ref{tab2} are the characteristic energy 
quanta associated with the rotational motion of the diatomic system. They 
are comparable with dispersion energies around distances of $R=100$, but
become dominant at interatomic separations of $1000$ Bohr radii and more.
As is well known, at ultracold temperatures only rotational s-waves remain
unaffected by the long-range rotational barrier and can probe interatomic 
interaction properties. In the presence of an external magnetic field, 
another energy scale must be taken into consideration: the Zeeman splitting,
which is of order $10^{-7}$ Hartree for a magnetic field strength of 
$100$ Gauss. Hence, typical laboratory fields are unable to distort the
spin-orbit coupling pattern in $^{88}$Sr, and our assumption of uncoupled
fine-structure states remains valid.

\begin{figure}
\includegraphics[width=3in,origin=c,angle=0]{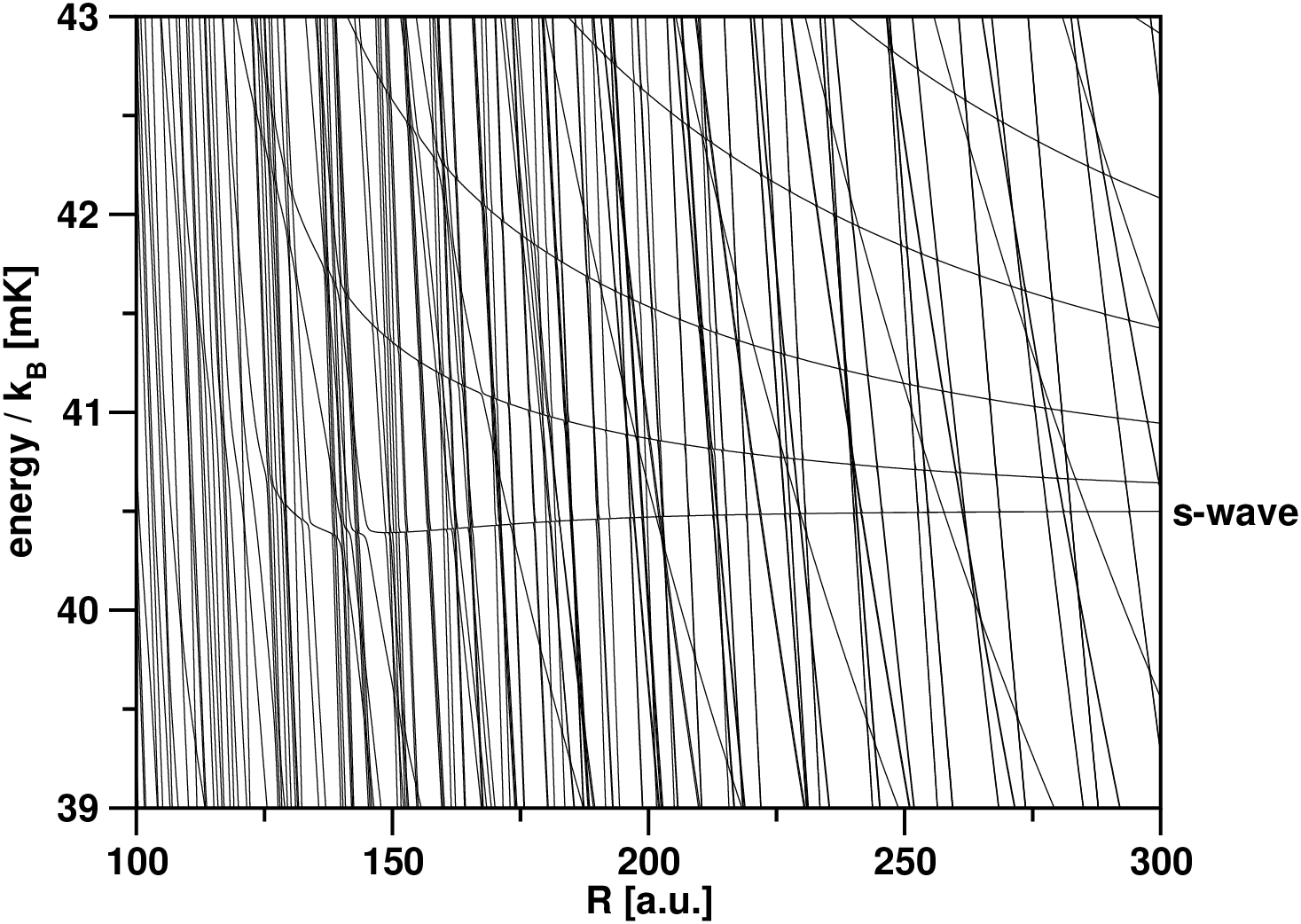}
\caption[]{Adiabatic potential energy curves of two metastable
strontium atoms
($\left(5{\mathrm{s}}5{\mathrm{p}}\right)$~$\phantom{}^{3}P\phantom{}_{2}$)
in a magnetic field of $100$~G, obtained by diagonalizing the matrix ${\bm V}$
defined in Eq.~(\ref{s4s2e1}). The total angular momentum projection quantum
number, $M_{\mathrm{tot}}$, is $+4$. The marked curve correlates at large
interatomic separations to two fully polarized strontium atoms (atomic
projection quantum number $m = +2$) in a rotational s-wave state ($L = 0$).
A basis describing molecular rotations in the laboratory frame with elements
up to $L = 40$ has been employed in the calculation.}
\label{fig1}
\end{figure}

\begin{figure}
\includegraphics[width=3in,origin=c,angle=0]{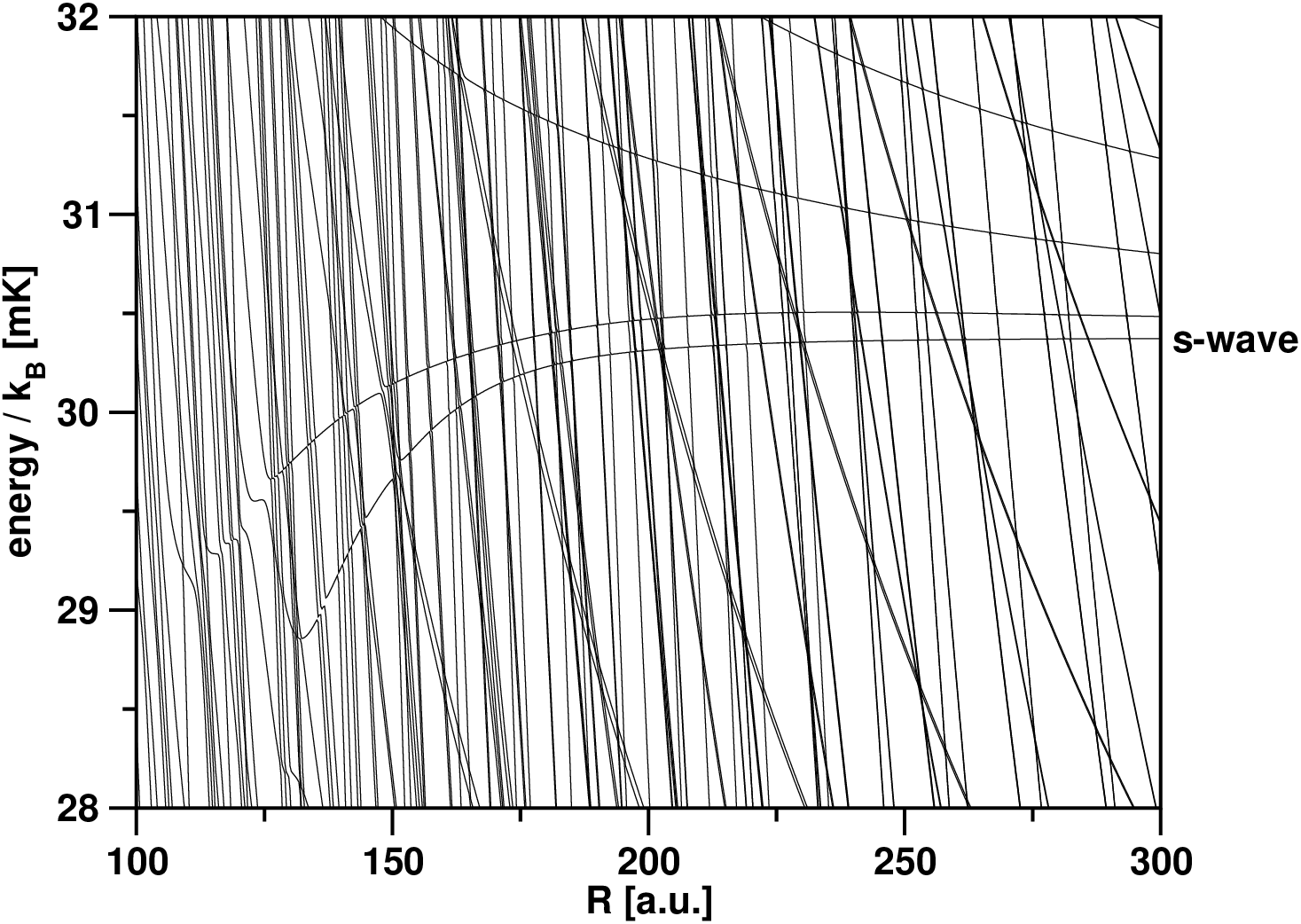}
\caption[]{Adiabatic potential energy curves of two metastable
strontium atoms
($\left(5{\mathrm{s}}5{\mathrm{p}}\right)$~$\phantom{}^{3}P\phantom{}_{2}$)
in a magnetic field of $100$~G. In this case $M_{\mathrm{tot}} = +3$.
The energy region shown corresponds to the Zeeman manifold deriving from
the situation that one atom is in an $m = +2$ state and the second one in an
$m = +1$ state. The atomic projection quantum numbers lose their validity
below distances of a few hundred Bohr radii, where atom-atom interactions
become important.}
\label{fig2}
\end{figure}

\begin{figure}
\includegraphics[width=3in,origin=c,angle=0]{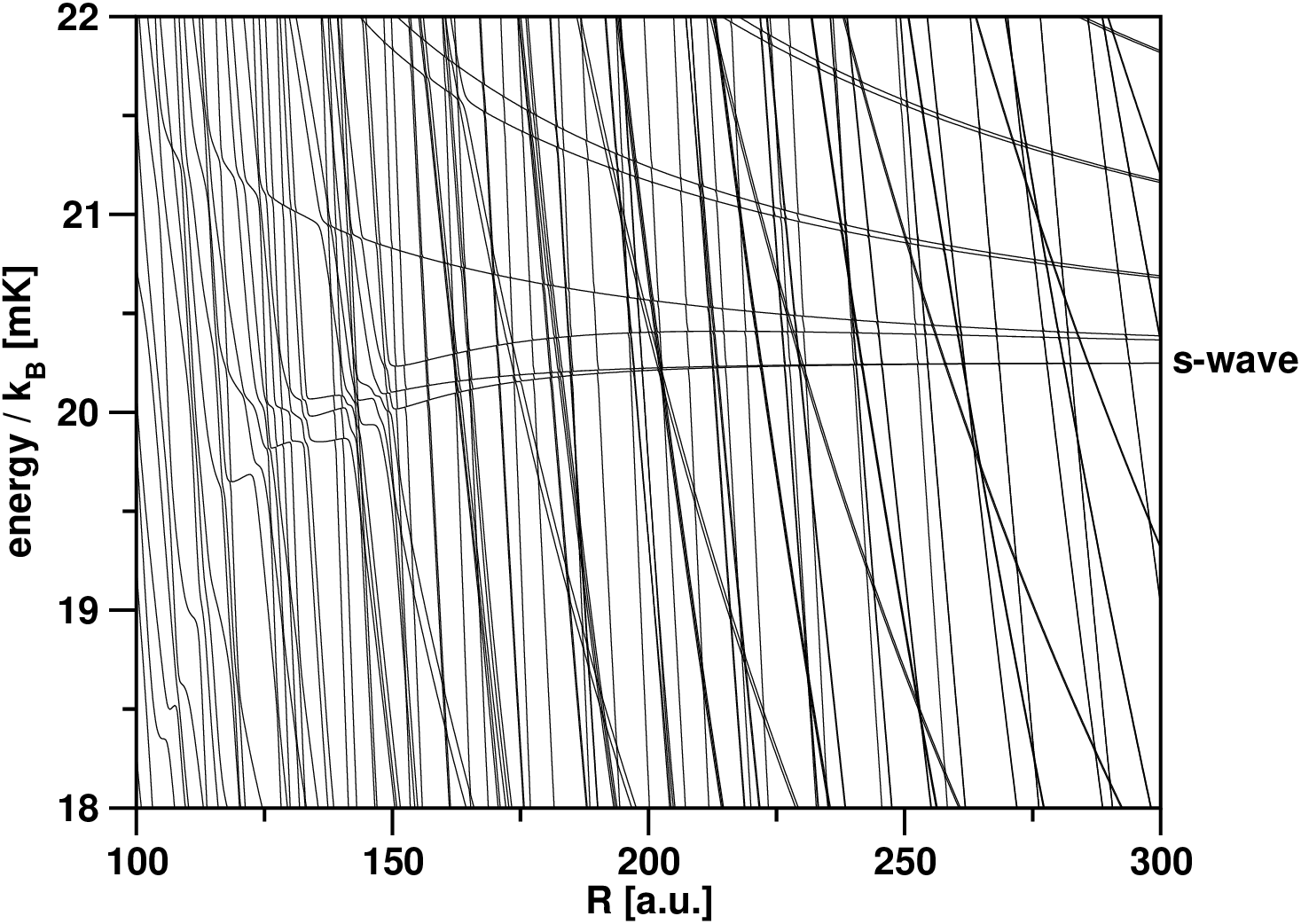}
\caption[]{Adiabatic potential energy curves of two metastable
strontium atoms
($\left(5{\mathrm{s}}5{\mathrm{p}}\right)$~$\phantom{}^{3}P\phantom{}_{2}$)
in a magnetic field of $100$~G, calculated setting $M_{\mathrm{tot}}$ to $+2$.
Here, the focus is on the Zeeman manifold that correlates to two atoms with
$m = +1$.}
\label{fig3}
\end{figure}

Figures~\ref{fig1}, \ref{fig2}, and \ref{fig3} display adiabatic 
potential energy curves of two metastable strontium atoms exposed to a 
magnetic field of $100$~G. The curves were calculated by diagonalizing,
as a function of $R$, the matrix
\begin{widetext}
\begin{eqnarray}
\label{s4s2e1}
\left<J_{\mathrm{tot}},M_{\mathrm{tot}},L,J\right|
\frac{{\bm L}^2}{2 m_{\mathrm{red}} R^{2}} + H_{\mathrm{I}}
-({\bm \mu}^{(1)} + {\bm \mu}^{(2)})_{0} {\cal B}
\left|\right.J^{''}_{\mathrm{tot}},M^{''}_{\mathrm{tot}},
L^{''},J^{''}\left.\right> \\
=: \left<J_{\mathrm{tot}},M_{\mathrm{tot}},L,J\right|
V\left|\right.J^{''}_{\mathrm{tot}},M^{''}_{\mathrm{tot}},
L^{''},J^{''}\left.\right>, \nonumber
\end{eqnarray}
\end{widetext}
i.e., the sum of the matrices of the rotational kinetic energy operator
(Eq.~(\ref{s3s2e2})), the interatomic interaction operator 
(Eqs.~(\ref{s3s1e7}) and (\ref{s3s1e8})), and the Zeeman operator
(Eq.~(\ref{s3s2e4})). Exploiting the symmetry properties of $V$, 
we selected a basis characterized by even rotational and even electronic
angular momentum quantum numbers. In our numerical studies we included
rotational quantum numbers up to $L_{\mathrm{max}} = 40$, which gave converged
results.

Of immediate interest for magnetic trap experiments are low-field
seeking atomic states. For $j = 2$ these are the ones with projection quantum
numbers $m=+1$ and $+2$. If both atoms are fully polarized ($m=+2$) and in
a rotational s-wave state (this is associated with a molecular eigenstate only
at large interatomic separations), then the total angular momentum projection 
quantum number, $M_{\mathrm{tot}}$, is $+4$. $M_{\mathrm{tot}}$ is conserved 
at all stages of a collision between the two atoms. The energetically highest 
Zeeman manifold with $M_{\mathrm{tot}} = +4$ is depicted in Fig.~\ref{fig1}. 
The potential energy curve correlating to the rotational s-wave is indicated in
the figure. The behavior of the s-wave energy differs significantly from that 
of the curves deriving from higher rotational states, which are clearly 
repulsive. The s-wave curve is attractive at distances larger than $150$ Bohr
radii and becomes repulsive at smaller radii. Near the minimum
of the resulting potential well there are several avoided crossings, which 
lead to inelastic losses through nonadiabatic transitions to lower lying 
Zeeman states. Because of the steepness of the diabatically crossing channels,
however, we do not expect severe losses and neglect them in this paper.

If one of the colliding atoms has $m= +2$ and the other one $m= +1$ at large
distances, then $M_{\mathrm{tot}} = +3$ assuming $L = 0$. The final combination
of two low-field seeking atoms---$m= +1$ for both atoms---corresponds to 
$M_{\mathrm{tot}} = +2$, again implying an s-wave collision. The two cases,
$M_{\mathrm{tot}} = +3$ and $M_{\mathrm{tot}} = +2$, are shown in 
Figs.~\ref{fig2} and \ref{fig3}, respectively. The most dramatic difference
between the cases $M_{\mathrm{tot}} = +4$ (Fig.~\ref{fig1}) and 
$M_{\mathrm{tot}} = +3$ (Fig.~\ref{fig2}) is the strength of the attractive
potential an s-wave experiences: For $M_{\mathrm{tot}} = +3$, this is 
stronger by about one order of magnitude. In this Zeeman manifold the 
couplings are so strong that even a state correlating to a higher partial 
wave is subject to strong attractive forces. In Fig.~\ref{fig3}, 
$M_{\mathrm{tot}} = +2$, the potential energy curves are even more 
complicated and suggest interesting cold-collision dynamics. There is a 
double degeneracy in this energy region at large interatomic separations,
since two atoms with $m = +1$ have the same energy as one atom with
$m = +2$ plus a second atom with $m = 0$. The latter atom cannot be 
magnetically trapped.

\begin{figure}
\includegraphics[width=3in,origin=c,angle=0]{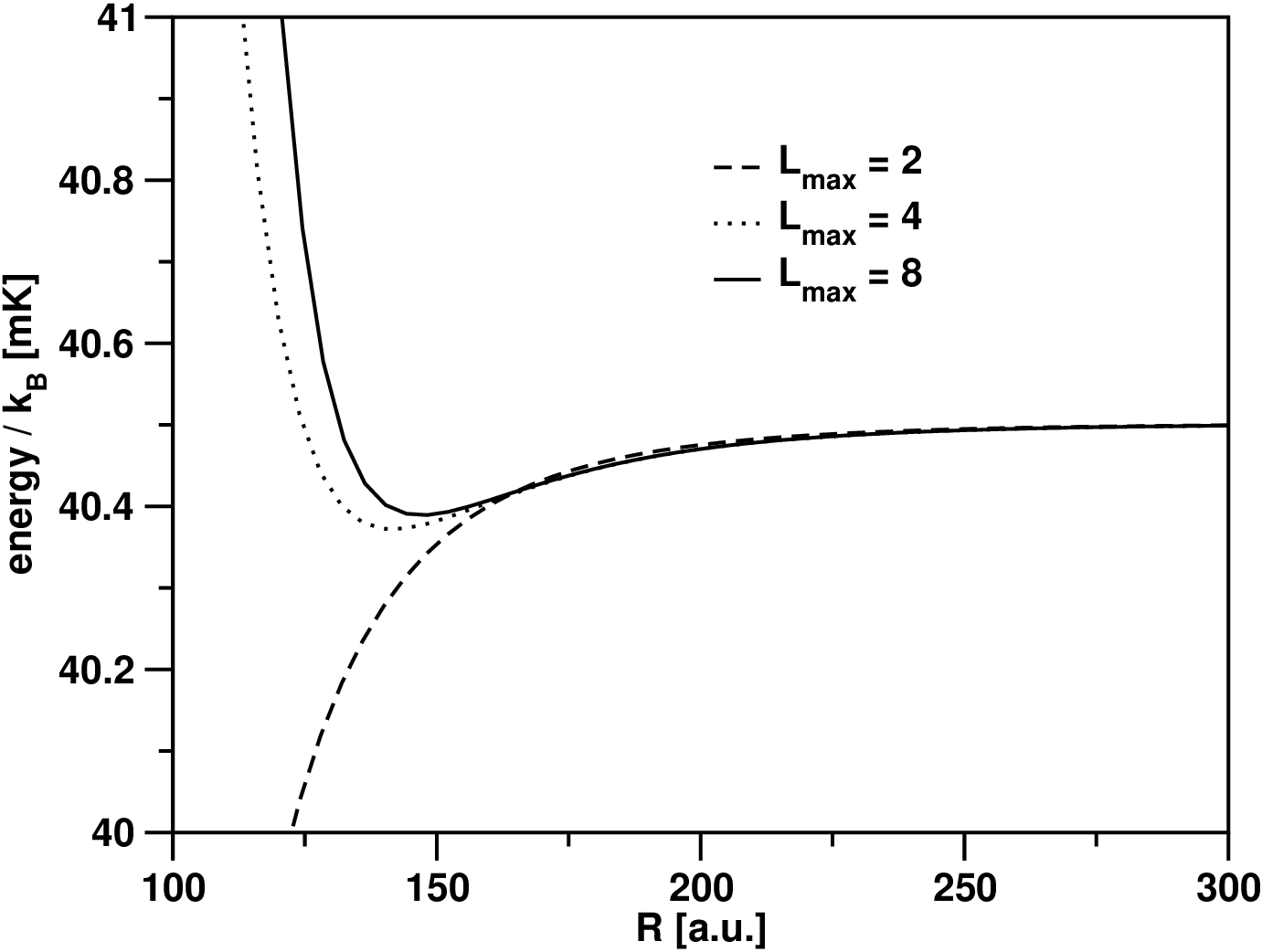}
\caption[]{Dependence of the potential energy curve of the ``s-wave''
in Fig.~\ref{fig1} on the largest molecular rotation quantum number,
$L_{\mathrm{max}}$, taken into account in the construction of the matrix
${\bm V}$ (Eq.~(\ref{s4s2e1})). Only if g-waves are included does a potential
well appear. The cause of the pronounced admixture of higher partial waves
is the relatively strong, anisotropic quadrupole-quadrupole interaction.
At large $R$, the curves converge to $40.502$~mK, corresponding to the Zeeman
energy of two metastable strontium atoms with $m = +2$ in a magnetic field
of $100$~G.}
\label{fig4}
\end{figure}

In the following we are going to concentrate on the long-range potential 
well in the energetically highest Zeeman manifold with 
$M_{\mathrm{tot}} = +4$ (Fig.~\ref{fig1}). Molecular rotations are found to 
play a key role in its origin. We have calculated adiabatic 
potential energy curves for three different rotational basis set sizes and
extracted those curves that correlate to two fully polarized strontium atoms
in a rotational s-wave state. The results are shown in Fig.~\ref{fig4}.
If in addition to s-waves only d-waves are 
included---$L_{\mathrm{max}} = 2$---the potential energy curve is purely 
attractive (the same holds true if a pure s-wave basis is used). Only as soon
as g-waves are taken into account ($L_{\mathrm{max}} = 4$) is there an 
effective repulsion below $R = 150$, thus leading to the emergence of a 
potential well minimum at large $R$. Nevertheless, the structure of the curve 
is still not converged. The potential energy curve obtained using 
$L_{\mathrm{max}} = 8$ illustrates the need for i- and k-waves. Larger 
rotational basis sets make it difficult to isolate a smooth potential energy 
curve due to the appearance of pronounced avoided crossings (see 
Fig.~\ref{fig1}). However, the structure of the effective diabatic curve 
correlating at large distances to an s-wave is essentially the one found for 
$L_{\mathrm{max}} = 8$.

Deeper insight into the formation of the long-range potential well and the
role played by molecular rotations can be gained by resorting to a 
perturbative approach. For that purpose we take the diagonal of the matrix
${\bm V}$ defined in Eq.~(\ref{s4s2e1}) to represent the unperturbed problem;
the perturbation is given by the off-diagonal elements. At large interatomic
separations the state vector of two fully polarized metastable strontium atoms
in a rotational s-wave state reads
\begin{eqnarray}
\label{s4s2e2}
\left|J_{\mathrm{tot}},M_{\mathrm{tot}},L,J\right>
& = & \left|4,4,0,4\right> \\
& =:& \left|\Phi_{0}\right>. \nonumber 
\end{eqnarray}
Up to first order its energy is 
\begin{equation}
\label{s4s2e3}
\left<\Phi_{0}|V|\Phi_{0}\right> = -\frac{C_{6,0}}{R^{6}} - 2
\left<\mu\right> {\cal B},
\end{equation}
where $C_{6,0} = 5.33 \times 10^{3}$~a.u. Hence we see that the long-range 
attraction in the s-wave channel is entirely due to dispersion forces;
more precisely, it derives from the scalar tensor ${\bm T}_{0}$ (see 
Eqs.~(\ref{s2s2e13}), (\ref{s3s1e8}), and (\ref{s3s1e9})).

\begin{figure}
\includegraphics[width=3in,origin=c,angle=0]{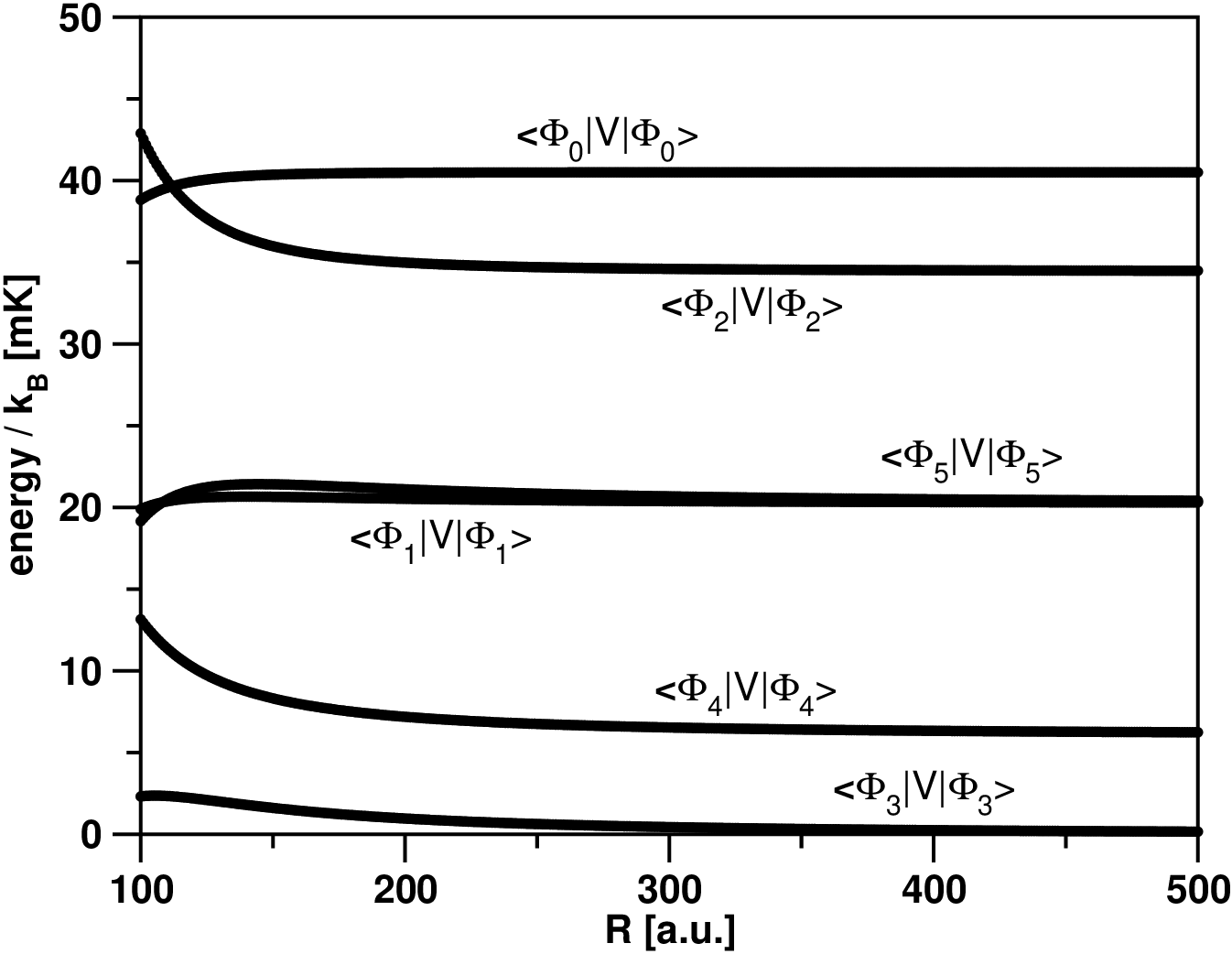}
\caption[]{Energy expectation values of the s-wave $\left|\Phi_{0}\right>$
(Eq.~(\ref{s4s2e2})) and the five basis states that can couple directly to it
(Eq.~(\ref{s4s2e4})). $\left|\Phi_{1}\right>$ and $\left|\Phi_{2}\right>$ are
d-waves; $\left|\Phi_{3}\right>$, $\left|\Phi_{4}\right>$, and
$\left|\Phi_{5}\right>$ are g-waves.}
\label{fig5}
\end{figure}

In order to evaluate the second-order correction, all basis vectors must be
determined that can couple directly to $\left|\Phi_{0}\right>$. Since 
$\left|\Phi_{0}\right>$ is an eigenvector of both the rotational kinetic
energy operator and the Zeeman operator, the coupling must be mediated by the
anisotropic interatomic interaction. $H_{\mathrm{I}}$ preserves  
$J_{\mathrm{tot}}$ and couples $\left|\Phi_{0}\right>$ to the basis vectors
\begin{eqnarray}
\label{s4s2e4}
\left|\Phi_{1}\right> & := & \left|4,4,2,2\right>, \\
\left|\Phi_{2}\right> & := & \left|4,4,2,4\right>, \nonumber \\
\left|\Phi_{3}\right> & := & \left|4,4,4,0\right>, \nonumber \\
\left|\Phi_{4}\right> & := & \left|4,4,4,2\right>, \nonumber \\
\left|\Phi_{5}\right> & := & \left|4,4,4,4\right>. \nonumber 
\end{eqnarray}
The first two are rotational d-waves, the other three are g-waves.
Their first-order energies, together with that of $\left|\Phi_{0}\right>$,
are depicted in Fig.~\ref{fig5}. Note that the basis vectors
$\left|J_{\mathrm{tot}},M_{\mathrm{tot}},L,J\right>$ are eigenvectors of
the Zeeman operator only if $L$ or $J$ vanishes. Therefore, the 
long-range limits of the first-order energies do not in general agree with
the Zeeman energies of two noninteracting atoms. In view of Fig.~\ref{fig5}
it is tempting to conclude that the strong repulsion that gives rise to 
the molecular potential well is due to the d-wave $\left|\Phi_{2}\right>$,
whose energy comes close to and eventually crosses the energy of the s-wave
channel.

\begin{figure}
\includegraphics[width=3in,origin=c,angle=0]{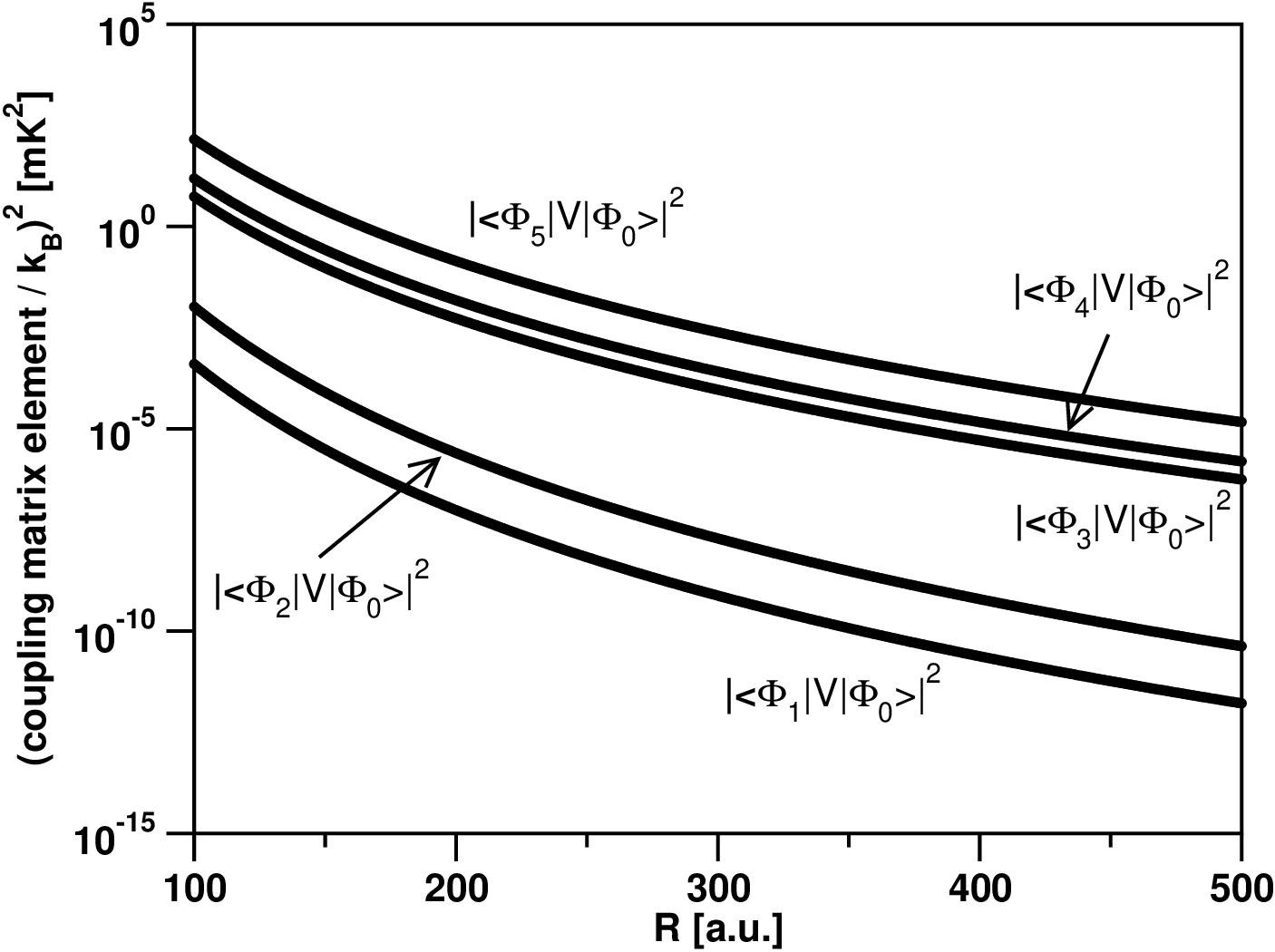}
\caption[]{Coupling strengths of the basis states $\left|\Phi_{1}\right>$
through $\left|\Phi_{5}\right>$ (Eq.~(\ref{s4s2e4})) to the s-wave
$\left|\Phi_{0}\right>$ (Eq.~(\ref{s4s2e2})), plotted on a logarithmic scale.
The coupling increases rapidly as the interatomic distance is reduced. It is
strongest for the g-waves $\left|\Phi_{3}\right>$, $\left|\Phi_{4}\right>$,
and $\left|\Phi_{5}\right>$, which can couple to $\left|\Phi_{0}\right>$ via
electric quadrupole-quadrupole interaction. The coupling of
$\left|\Phi_{0}\right>$ to the d-waves $\left|\Phi_{1}\right>$ and
$\left|\Phi_{2}\right>$ is mediated by the comparatively weak dispersion
interaction.}
\label{fig6}
\end{figure}

\begin{figure}
\includegraphics[width=3in,origin=c,angle=0]{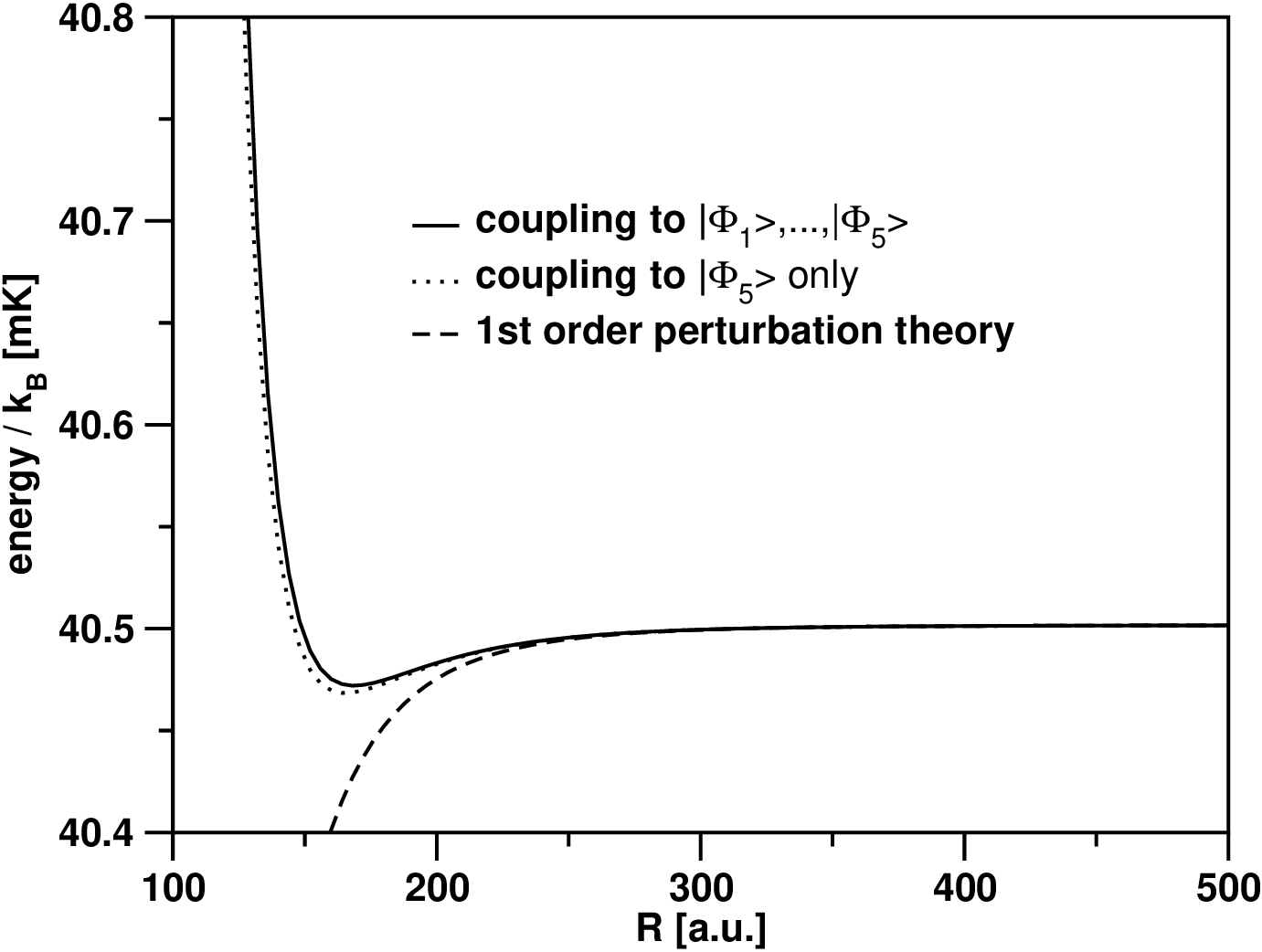}
\caption[]{In first-order perturbation theory, the potential energy curve
of the s-wave $\left|\Phi_{0}\right>$ is purely attractive
(Eq.~(\ref{s4s2e3})). Due to the strong coupling to the g-waves, in particular
$\left|\Phi_{5}\right>$, a potential well emerges. An approximate analytic
representation is given in Eq.~(\ref{s4s2e5}).}
\label{fig7}
\end{figure}

However, this is not true. In Fig.~\ref{fig6} we have plotted the modulus 
squared of the coupling matrix elements 
$\left<\Phi_{i}\right|V\left|\Phi_{0}\right>$, $i=1,...,5$. Note the use of
a logarithmic scale along the ordinate. The coupling of the s-wave to the 
g-waves $\left|\Phi_{3}\right>$, $\left|\Phi_{4}\right>$, and 
$\left|\Phi_{5}\right>$ turns out to be stronger by many orders of magnitude
than the coupling to the d-waves $\left|\Phi_{1}\right>$ and 
$\left|\Phi_{2}\right>$. The reason is the tensorial structure of 
$H_{\mathrm{I}}$ (Eqs.~(\ref{s2s2e13}) and (\ref{s3s1e5})): The d-waves 
couple to $\left|\Phi_{0}\right>$ through the relatively weak second-rank
tensor ${\bm T}_{2}$; the g-waves, however, couple to the s-wave via the 
fourth-rank tensor ${\bm T}_{4}$, which is dominated by the strong electric 
quadrupole-quadrupole interaction. Because $\left|\Phi_{5}\right>$ is closest
in energy to $\left|\Phi_{0}\right>$ (Fig.~\ref{fig5}) and its coupling matrix
element is by far the greatest (Fig.~\ref{fig6}), it is likely that it 
accounts for most of the second-order correction to the s-wave energy. 
Figure~\ref{fig7} confirms this. The difference between the full
second-order energy and that due to $\left|\Phi_{5}\right>$ alone is 
relatively small. What is even more important is the difference to the 
first-order energy. In first order, the s-wave experiences pure attraction by 
scalar dispersion forces. As a consequence of strong quadrupole-quadrupole 
coupling to g-waves associated with energetically lower Zeeman manifolds,
a steep repulsive potential wall appears. This requires that coupling is 
introduced at least to second order in perturbation theory. 

The simple perturbative approach does not provide quantitative agreement
with our numerical results, but it evidently captures the basic physics 
underlying the existence of the long-range potential well. Moreover, it
predicts that the properties of the potential well depend on
${\cal B}$, the magnetic-field strength. The energy difference between
$\left|\Phi_{0}\right>$ and the most important g-wave, $\left|\Phi_{5}\right>$,
is approximately independent of interatomic distance (Fig.~\ref{fig5}).
This can be exploited for a compact approximate representation of the 
second-order energy of $\left|\Phi_{0}\right>$, valid at magnetic fields 
of about $100$~G and higher:
\begin{eqnarray}
\label{s4s2e5}
E & = & - 2 \left<\mu\right> {\cal B} \\
& & -\frac{C_{6,0}}{R^{6}} - \frac{1}{\left<\mu\right> {\cal B}}
\frac{\kappa}{R^{10}}. \nonumber
\end{eqnarray}
The quadrupole-quadrupole coupling parameter $\kappa$ is 
$1.44 \times 10^{5}$~a.u. The location, with respect to $R$, of the potential 
well minimum described by Eq.~(\ref{s4s2e5}) scales as 
${\cal B}^{-1/4}$, while the well depth scales as ${\cal B}^{3/2}$. 
As the magnetic-field strength is increased,
the effect of electric quadrupole-quadrupole coupling at a given distance $R$ 
diminishes. In order to compensate for the larger Zeeman splittings, it is 
necessary to go to shorter distances. Along the way the dispersion attraction 
grows stronger, until it is eventually overwhelmed by repulsion induced by 
electric quadrupole-quadrupole interaction. The potential well therefore gets 
deeper and its minimum is shifted to smaller interatomic distances as 
${\cal B}$ is ramped up.

The tunability of the molecular potential well could prove useful for 
controlling the properties of metastable, magnetically trapped $^{88}$Sr.
The most important quantity in this context is the scattering length $a$
\cite{Saku94}, which characterizes the effective interaction between ultracold
collision partners \cite{PeSm02}. A positive $a$ describes effective repulsion;
negative $a$ implies attraction. The strength of the interaction is 
governed by $|a|$.

\begin{figure}
\includegraphics[width=3in,origin=c,angle=0]{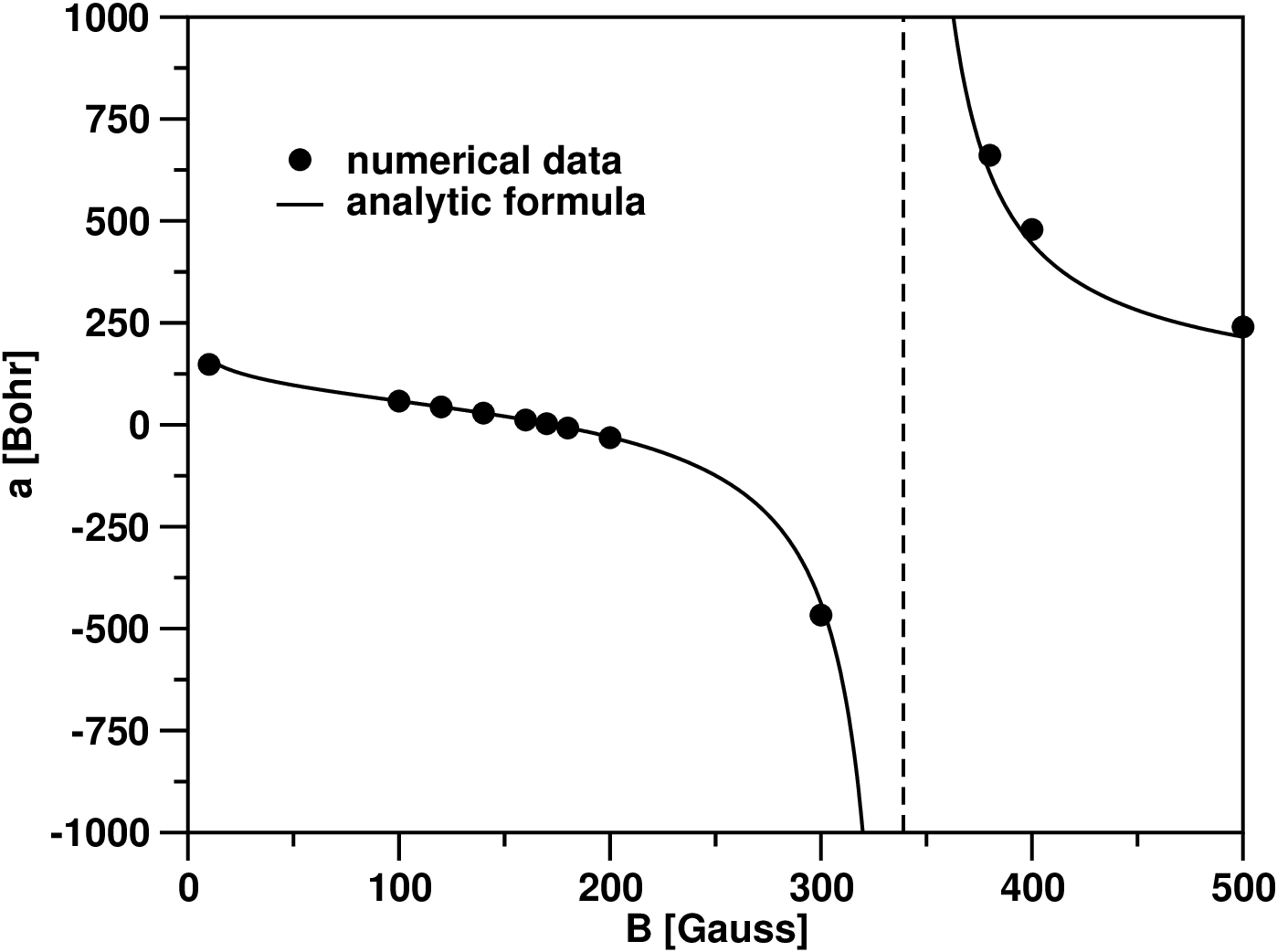}
\caption[]{Scattering length, $a$, in the long-range potential well
calculated as a function of the magnetic-field strength, ${\cal B}$.
The dots represent numerical data obtained by direct integration of the
one-dimensional Schr\"{o}dinger equation. The solid line is based on the
analytic formula in Eq.~(\ref{s4s2e6}) (${\cal B}_{0} = 173$~G and 
$a'({\cal B}_{0}) = -0.98/$G).}
\label{fig8}
\end{figure}

We have therefore calculated the scattering length in the long-range potential
well as a function of magnetic-field strength, ${\cal B}$. Our results are 
displayed in Fig.~\ref{fig8}. For each ${\cal B}$ value, we determined, in a 
procedure similar to the one that led to Fig.~\ref{fig4}, the maximum $L$ 
that gave a smooth potential energy curve compatible with the respective 
result for $L_{\mathrm{max}} = 40$, but which no longer exhibits pronounced
avoided crossings. We then numerically integrated the one-dimensional
Schr\"{o}dinger equation with the selected potential energy curve and for
asymptotically vanishing kinetic energy from $R_{\mathrm{min}} = 40$ to
$R_{\mathrm{max}} = 1940$~a.u. The scattering length was obtained from the 
logarithmic derivative of the computed wavefunction at $R = R_{\mathrm{max}}$.
We verified that our results are converged with respect to variations of the 
integration limits $R_{\mathrm{min}}$ and $R_{\mathrm{max}}$.

We have restricted the lowest magnetic-field strength in Fig.~\ref{fig8}
to ${\cal B} = 10$~G, because below that the potential well starts to 
disappear. At ${\cal B} = 0$, the long-range potential is purely attractive,
and additional knowledge about the short-range part of the potential would be 
necessary in order to make quantitative predictions. The strongest field we 
have considered is ${\cal B} = 500$~G. The left classical 
turning point at this field strength is already down to about $R= 80$. Still 
higher field strengths would again maneuver us into a regime where our theory 
probably becomes unreliable (see the discussion at the beginning of this 
subsection). However, between $10$ and $500$ Gauss---and within the 
approximation of pure elastic scattering---our scattering lengths may be 
regarded as accurate to within about $10\%$.

There are two important features about the ${\cal B}$-dependence. First, the 
scattering length vanishes at ${\cal B}_{0} = 173$~G. The slope at this 
point is $a'({\cal B}_{0}) = -0.98/$G. Second, $a$ diverges at the resonance
field strength ${\cal B}_{\mathrm{res}} = 339$~G, which signals the appearance
of the first bound state in the long-range potential well. In the vicinity 
of both ${\cal B}_{0}$ and ${\cal B}_{\mathrm{res}}$ the effective interatomic
interaction can be switched between attraction and repulsion. Near 
${\cal B}_{\mathrm{res}}$, the ability to tune $a$ over an extensive range
may be particularly advantageous.

Interestingly, ${\cal B}_{\mathrm{res}}$ and ${\cal B}_{0}$ are not 
independent of one another. We can derive an approximate analytic formula for 
the scattering length as a function of magnetic-field strength:
\begin{eqnarray}
\label{s4s2e6}
a({\cal B}) & = & -\frac{8}{5\pi} a'({\cal B}_{0}) {\cal B}_{0} \\
& & \times \left[1 - 
\tan{\left(\frac{5\pi}{8}\sqrt{{\cal B}/{\cal B}_{0}} - \frac{3\pi}{8}\right)}
\right],
\nonumber
\end{eqnarray}
which depends only on the parameters ${\cal B}_{0}$ and $a'({\cal B}_{0})$.
Equation~(\ref{s4s2e6}) is based on a semiclassical analysis by Gribakin and
Flambaum \cite{GrFl93}. We have employed their analytic result for a 
hard-core plus a $1/R^{n}$-potential. In our case, the attraction is 
basically due to pure dispersion interaction and the repulsion is so strong 
that replacing it by a hard core is not such a bad approximation. In accordance
with the discussion following Eq.~(\ref{s4s2e5}), we assumed the onset of the
hard core to scale as ${\cal B}^{-1/4}$. The first resonance in 
Eq.~(\ref{s4s2e6}) appears when the argument of the $\tan$-function equals
$\pi/2$, or ${\cal B}/{\cal B}_{0} = (7/5)^{2} = 1.96$. This is in excellent 
agreement with our numerical data: 
${\cal B}_{\mathrm{res}}/{\cal B}_{0} = 339/173 = 1.96$. The scattering 
length as predicted by Eq.~(\ref{s4s2e6}), using the parameters
${\cal B}_{0} = 173$~G and $a'({\cal B}_{0}) = -0.98/$G estimated from our 
numerical data, is also plotted in Fig.~\ref{fig8}. The analytic formula 
reproduces the magnetic-field dependence of the scattering length remarkably
well.

\section{Conclusion}
\label{sec5}

Potential wells associated with ordinary molecular bonds are typically a
few electron volts, or $10^{4}$ Kelvin, deep; the minimum of the potential 
energy is found at interatomic distances of the order of one Bohr radius.
Electronic interactions at such short distances are so strong that the energy
separation between the ground and excited electronic states dissociating to
the same atomic configuration is very large in comparison to rotational 
energies. In a simplified picture we might envisage the strong chemical bond
as forcing the atoms into a fixed orientation with respect to the interatomic
axis. Molecular rotation leaves this rigid arrangement virtually unaffected.

In fact, however, the long-range physics of metastable alkaline-earth atoms is
fundamentally different from that simplistic picture. If the atoms are 
separated by a sufficiently large distance, they have constant angular momenta.
As the atoms approach one another, they experience anisotropic forces that 
modify the molecular rotational motion as well as the relative orientation of 
the nonspherical atoms. At interatomic distances of a few hundred Bohr radii, 
the energy scales of interatomic interactions and molecular rotations are 
comparable, and the coupling between atomic and molecular angular momentum 
turns out to be rather efficient.

The tensorial analysis presented in this paper puts the coupling mechanism
into a particularly clear and useful form. Our formulation makes maximum use
of symmetries, which allows a compact matrix representation of the long-range
Hamiltonian. It will enable systematic multichannel scattering calculations
needed to investigate the role played by inelastic collision processes.

In this study we have numerically diagonalized the Hamiltonian matrix for two 
metastable strontium atoms and obtained adiabatic potential energy curves,
the adiabaticity referring only to the {\em distance} coordinate $R$. The 
curves are in general very complicated. We have therefore focussed on one 
curve---associated with an s-wave channel and with the atoms in an 
experimentally attractive low-field seeking state---that exhibits a novel 
type of long-range potential well. This well arises from a fascinating 
interplay between the pure $1/R^{6}$-attraction in the s-wave channel and the 
strong quadrupole-quadrupole coupling to rotational g-waves attached to lower
lying Zeeman manifolds. Recently, Avdeenkov and Bohn discovered that the 
anisotropic forces between polar OH molecules in an electrostatic field
lead to a similar phenomenon \cite{AvBo02}.

The scattering length in the long-range potential well in metastable strontium 
can be tuned by varying the external magnetic-field strength. We find a 
resonance at $339$~G, which differs quantitatively from the prediction by 
Derevianko {\em et al.} \cite{DePo03}, who estimate the resonance location to 
lie near $1000$~G. We believe that the present calculations should improve 
upon the accuracy of the more approximate treatment of Ref.~\cite{DePo03}.

Finally, the nature of the resonance deserves attention. It indicates the 
emergence of the first vibrational bound state as the long-range potential 
well undergoes a controlled deformation. This process must be contrasted with
the physics underlying Feshbach resonances \cite{InAn98,CoFr98,RoCl98,CoCl00},
where the scattering wave in the entrance channel is brought into resonance
with a vibrational bound state in a potential well correlating to an 
energetically higher channel. The potential energy curves themselves do not
change their qualitative appearance. The strong anisotropic interactions in 
cold gases of metastable alkaline-earth atoms open an entirely new route to 
shaping potential energy curves and creating temporary molecules with bond 
lengths of the order of a hundred Bohr radii.

\acknowledgments
We would like to thank Kevin Christ for contributing his one-electron 
finite-element code. We thank Hossein Sadeghpour and Andrei Derevianko for
discussions. R.S. gratefully acknowledges financial support by the 
Emmy Noether program of the German Research Foundation (DFG). This work
was supported in part by the Department of Energy, Office of Science.

\end{document}